\documentclass[iop,numberedappendix,useAMS,usenatbib]{emulateapj}
\usepackage{times}
\usepackage{apjfonts}
\usepackage{amssymb}
\usepackage{epsfig}

\newcommand{\Enzo}{\texttt{Enzo}}

\newcommand{\hif}{\textit{HIFLUGCS~}}

\def\M{{\mathcal{M}}}
\def\deg{{^{\circ}}}
\begin{document}

\shorttitle{How Much Can We Learn From A Merging Cold Front Cluster}
\shortauthors{Datta et al.}

\title{How Much Can We Learn From A Merging Cold Front Cluster? : Insights from X-Ray Temperature and Radio Maps of Abell 3667} 

\author{Abhirup Datta\altaffilmark{1,2,5}, David E. Schenck\altaffilmark{2}, Jack O. Burns\altaffilmark{2,5}, Samuel W. Skillman\altaffilmark{2,3} and Eric J. Hallman\altaffilmark{2,4}} 

\altaffiltext{1}{NASA Post-Doctoral Fellow}
\altaffiltext{2}{Center for Astrophysics and Space Astronomy, Department of Astrophysical and Planetary Science, University 
of Colorado, Boulder, C0 80309,USA}
\altaffiltext{3}{DOE Computational Science Graduate Fellow}
\altaffiltext{4}{Tech-X Corporation, Boulder, CO 80303}
\altaffiltext{5}{Lunar University Network for Astrophysics Research (LUNAR), NASA Lunar Science Institute, NASA Ames Research Center, Moffett Field, CA 94089}

\email{Abhirup.Datta@colorado.edu}

\begin{abstract}
The galaxy cluster Abell 3667 is an ideal laboratory to study the plasma processes in the intracluster medium (ICM). High resolution {\it Chandra} X-ray observations show a cold front in Abell 3667. At radio wavelengths, Abell 3667 reveals a double radio-relic feature in the outskirts of the cluster. These suggest multiple merger events in this cluster. In this paper, we analyze the substantial archival X-ray observations of Abell 3667 from {\it Chandra} X-ray Observatory and compare these with existing radio observations as well as state-of-the-art AMR (Adaptive Mesh Refinement) MHD cosmological simulations using \Enzo. We have used two temperature map making techniques, Weighted Voronoi Tessellation and Adaptive Circular Binning, to produce the high resolution and largest field-of-view temperature maps of Abell 3667. These high fidelity temperature maps allow us to study the X-ray shocks in the cluster using a new 2-dimensional shock-finding algorithm. We have also estimated the Mach numbers from the shocks inferred from previous ATCA radio observations. The combined shock statistics from the X-ray and radio data are in agreement with the shock statistics in a simulated MHD cluster. We have also studied the profiles of the thermodynamic properties across the cold front using $\sim 447$~ksec from the combined {\it Chandra} observations on Abell 3667. Our results show that the stability of the cold front in Abell 3667 can be attributed to the suppression of the thermal conduction across the cold front by a factor of $\sim 100-700$ compared to the classical Spitzer value.  
\end{abstract}


\section{Introduction}

In the hierarchical structure formation framework, clusters of galaxies are the largest approximately virialized objects in the universe and are ideal laboratories to study astrophysical plasma processes. Clusters are assembled through large and small mergers [see e.g. \citet{kravtsov12}]. During mergers cosmological shocks are driven into the intracluster medium (ICM). These shocks heat the ICM which is then detected in the soft X-ray regime through its thermal emission \citep{sarazin88,bohringer10}. In addition, these shocks accelerate non-thermal electrons and protons to relativistic speeds \citep{drury83,blandford87,skillman11,bruggen12}. The relativistic electrons have relatively short lifetimes ($10^8$ years) and emit synchrotron radiation which are detected at radio wavelengths. Mergers of galaxy clusters are the most energetic events in the universe since the Big Bang. Up to $10^{64}$~erg of gravitational potential energy is dispersed among the cluster constituents in a major merger \citep{ricker01}. 

\begin{figure*}[t!]
\centering
\epsfig{file=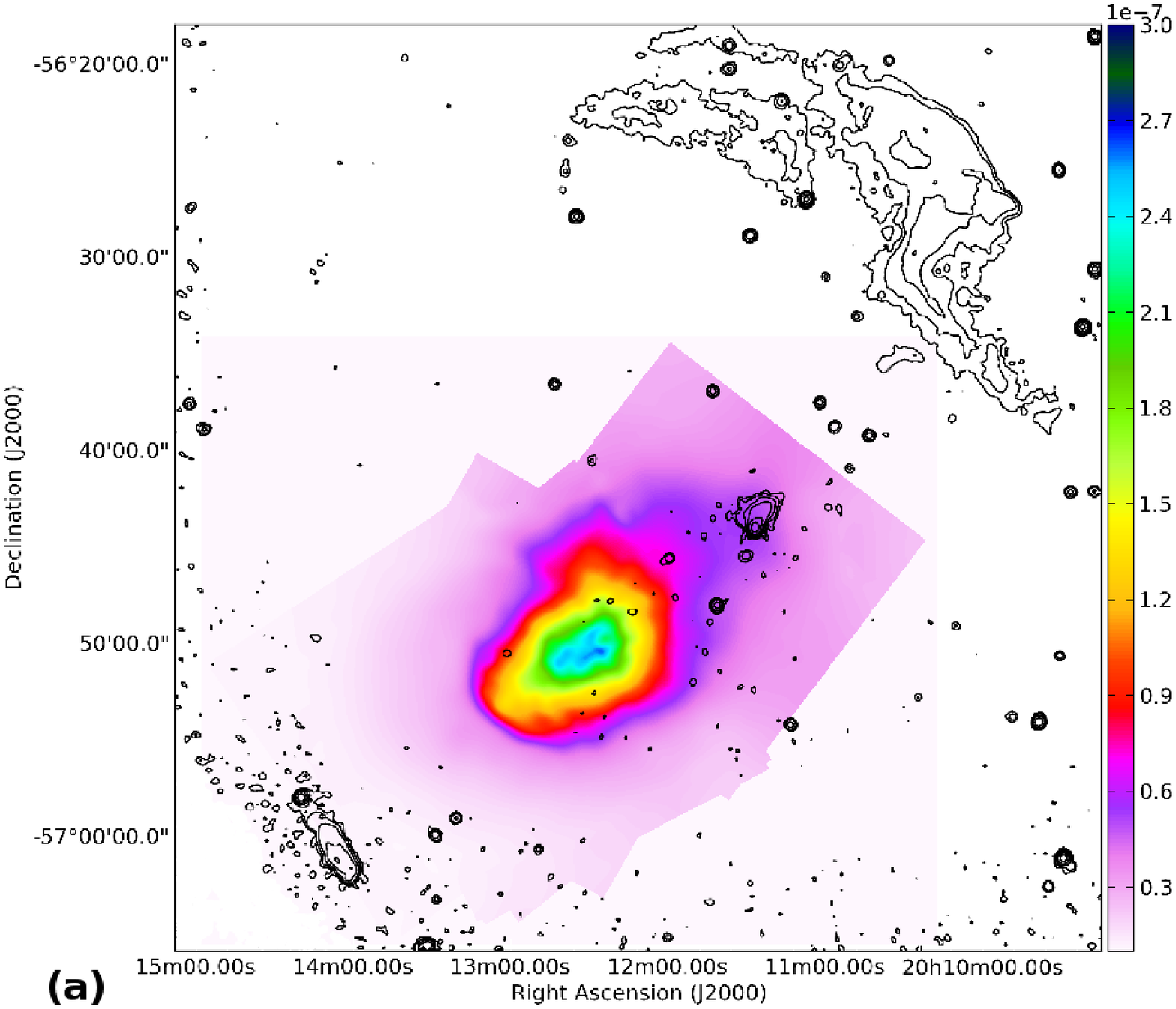,height=3.5truein}
\epsfig{file=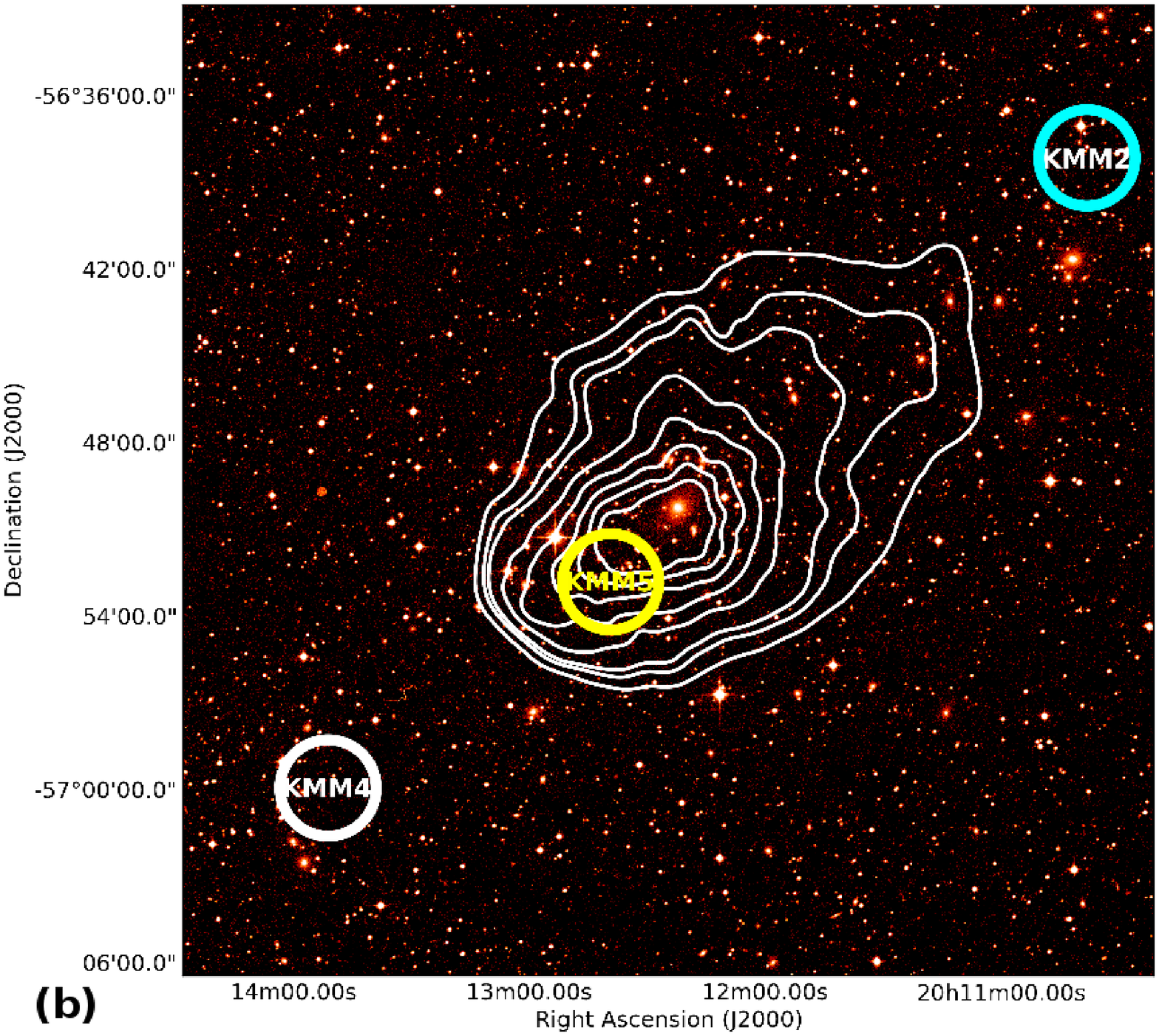,height=3.5truein}
\caption{{\bf (a)} The contours of radio map from the ATCA observations at 20cm wavelength \citep{huub97} are overlayed on the X-ray surface brightness map from the combined eight {\it Chandra} observations. The X-ray surface brightness units are ergs s$^{-1}$ cm$^{-2}$ arcmin$^{-2}$. {\bf (b)} The smoothed contours of the X-ray surface brightness map from the combined eight {\it Chandra} observations has been overlayed on the optical image from the SuperCOSMOS Sky Survey r-band data \citep{hambly01}. Also, we show the KMM regions defining the 3 major substructure in the A3667 cluster \citep{owers09a}. The exact locations of the South-East (in white: KMM4), main cluster (in yellow: KMM5) and the North-West substructure (in cyan: KMM2). Exact locations of these KMM regions are provided by Matt Owers (private communication).}
\label{fig:rgb}
\end{figure*}

In this paper, we analyze the very extensive X-ray observations of Abell 3667 from {\it Chandra} X-ray Observatory  (total effective exposure of 447 ksec) and compare these with existing radio observations as well as state of the art AMR (Adaptive Mesh Refinement) MHD cosmological simulations using \Enzo. Abell 3667 has been studied at X-ray, optical and radio wavelengths as a spectacular ongoing merger (see Figure~\ref{fig:rgb}). The observations carried out at all three wavelengths make A3667 an ideal laboratory to study the plasma processes in the ICM. Figure~\ref{fig:rgb}(a) shows the overlay of the radio contours at 1.4 GHz from ATCA (Australia Telescope compact array) on the {\it Chandra} X-ray surface brightness map. Figure~\ref{fig:rgb}(b) shows the overlay of the X-ray surface brightness contours on the optical image from SuperCOSMOS Sky Survey r-band data\footnote{http://www-wfau.roe.ac.uk/sss/pixel.html}. Abell 3667 is one of the \hif clusters at a redshift of $z=0.055$ \citep{sodre92}. It should be noted that the \hif or the HIghest X-ray FLUx Galaxy Cluster Sample is a complete sample of the X-ray brightest galaxy clusters \citep{reiprich02}. \citet{hudson10} studied the cool-core clusters in the \hif sample and classified A3667 as a weak cool-core (WCC) cluster. In other studies, Abell 3667 is classified as a non cool-core cluster \citep{rossetti10}. In the X-ray observations, the main feature seen in Abell 3667 is the cold front [see \citet{markevitch07} for a review of cold fronts in galaxy clusters]. Abell 3667 has an X-ray luminosity of $L_X(0.4-2.4~keV) = 5.1 \times 10^{44} erg~s^{-1}$ \citep{ebeling96}, $M_{200}=11.19\pm1.65~10^{14} h_{50}^{-1} M_\odot$ and global temperature $kT_x = 7\pm0.6$~keV \citep{markevitch98,vikhlinin01b,reiprich02}. This cluster is Abell richness class 2. The first X-ray observations with the {\it ROSAT} PSPC suggested an ongoing merger \citep{knopp96}. The sharpness of the discontinuity across the cold front in the ASCA images \citep{markevitch98} was initially presumed to be a shock front by \citet{markevitch99}. Later on, the analysis of a higher spatial and spectral resolution {\it Chandra} observation [obsid 513 in Table~\ref{tab:obs}] by \citet{vikhlinin01b} proposed this structure to be a cold front -- a contact discontinuity in A3667. This cluster has also been observed with {\it XMM Newton} by \citet{briel04}, \citet{lovisari09} and \citet{finoguenov10}. From the metallicity map covering the central parts of the cluster, \citet{lovisari09} suggested that about $65-80\%$ contribution from SN II (supernova type II) is required to obtain the observed metallicity abundance ratios near the center of A3667.

%------------------------ Figure:- 1 -----------------------------
%------------------------------------------------------------------

At radio wavelengths, A3667 shows a double radio-relic structure [see Figure~\ref{fig:rgb}(a) and \citet{huub97}] at 13 and 20 cm [see \citet{feretti12} for a review on radio relics]. The relics are located on either side of the cluster center at a distance of 1.8 Mpc. Recently, \citet{carretti13} observed A3667 with the Parkes telescope at 2.3 GHz and 3.3 GHz as part of the S-band Polarization All Sky Survey (S-PASS). They detected a radio bridge between the North-West relic and the center of the cluster. This bridge was also marginally detected at 843 MHz images from SUMMS (Sydney University Molonglo Sky Survey), but with no ICM emission from the central regions of the cluster \citep{carretti13}. These radio observations support the scenario of an ongoing merger in A3667 \citep{roettiger99}. So far, A3667 is the only double-relic system with a radio-bridge. This also suggests that the shocks are likely to propagate in the plane of the sky, thus removing the uncertainties about the projection effects \citep{carretti13}. \citet{finoguenov10} observed this cluster with XMM-Newton near the North-West radio relic and confirmed a shock from the X-ray temperature and density jumps at the relic location. 

At optical wavelengths, A3667 has been studied extensively by \citet{hollitt08} and \citet{owers09a}. \citet{owers09a} used multiobject spectroscopy with the 3.9 meter Anglo-Australian Telescope to search for substructure. They found that the cluster can be separated into two major substructures and the cold front appears to be directly related to the cluster merger activity. The substructure to the North-West (KMM2 in Figure~\ref{fig:rgb}(b)) is also consistent with the mass concentration found in the weak-lensing maps by \citet{joffre00}. The South-East substructure was tidally stripped from the main North-East substructure during its approach to the parent cluster core. 

\citet{akamatsu12} observed A3667 with Suzaku XIS (X-ray Imaging Spectrometer) instrument covering the region between the center and the North-West relic. These observations suggest that the gas is heated by a shock propagating away from the center to the outer region of the cluster. Previously, \citet{nakazawa09} used the Suzaku HXD (Hard X-ray Detector) to observe near the North-West relic region and inferred the magnetic field within the relic region to be $\gtrsim 1.6~\mu G$. This is consistent with the previous results from \citet{finoguenov10} using the XMM Newton where they deduced the magnetic field to be $\gtrsim 3~\mu G$. This lower limit on the magnetic field was obtained by comparing the limit on the inverse Compton X-ray emission with the measured radio synchrotron emission.

\citet{roettiger99} were able to produce the observed properties of the radio relics with the first 3-dimensional MHD/N-body simulations. The model of \citet{roettiger99} posits that A3667 has undergone a recent merger ($\lesssim 1$~Gyr) with a subcluster with total mass of $20\%$ of the primary cluster. This work deduced that the subcluster was moving from South-East to North-West and impacted the primary cluster slightly off-axis. The merger generated multiple shocks which are sites for diffusive shock acceleration \citep{drury83} of relativistic electrons. \citet{roettiger99} were able to explain the location and morphology of the radio relics with respect to the X-ray surface brightness. \citet{poole06} used a 3:1 mass ratio impact to reproduce the two shocks moving in opposite directions as well as producing a cold front. The interpretation of the substructure dynamics from \citet{owers09a} is consistent with the models given by \citet{roettiger99} and \citet{poole06}. \citet{kawahara08} analyzed the {\it Chandra} observations of A3667 to study the X-ray surface brightness fluctuations and found that it follows a lognormal distribution as seen in their synthetic clusters from cosmological hydrodynamic simulations. 

In this paper, we combined all archival {\it Chandra} observations for A3667 [see Table~\ref{tab:obs}] into one of the longest effective integration ($\sim 447$~ksec) of a galaxy cluster that has been heretofore available. All previous studies on A3667 using the {\it Chandra} observations were either based on limited exposure time \citep{vikhlinin01a,vikhlinin01b,vikhlinin02,vikhlinin03} or limited field-of-view around the cold front \citet{owers09b}. Combining 8 observations of A3667 from {\it Chandra} archive, we derive X-ray temperature maps covering the central $\sim 1$~Mpc region of the cluster using two different temperature map-making techniques.  In Section~\ref{sec:anal} of this paper, we discuss the X-ray data analysis pipeline that has been used to analyze the {\it Chandra} archival observations. Section~\ref{sec:temp} compares the two temperature map-making techniques and discusses the results. In Section~\ref{sec:sf}, we analyze the X-ray shocks in A3667 from the combined {\it Chandra} observations using a new automated shock-finder. In Section~\ref{sec:sim}, we compare the shock statistics from the X-ray data with the results from MHD AMR cosmological simulations. This is a first systematic study of the X-ray shocks in A3667 using {\it Chandra} observations. Since the {\it Chandra} observations are limited to the central $\sim 1$~Mpc of the cluster and do not extend to the location of the double relic ($\sim 1.8$~Mpc from the cluster center), in Section~\ref{sec:rad} we have used previous radio observations by \citet{huub97} to trace the shocks beyond the central $1$~Mpc region. In Section~\ref{sec:compXRS}, we validate the new shock-finder and compare the combined shock statistics from X-ray and radio observations of A3667 with an existing 3-dimensional shock-finder results from the cosmological MHD AMR simulations. Section~\ref{sec:cf} focuses on our analysis of the cold front in A3667. The higher signal-to-noise of the combined {\it Chandra} observations allow us to place a significantly tighter limit on the width of the cold front than the previous study by \citet{vikhlinin01a}. Profiles of thermodynamic quantities like density, temperature, pressure and entropy are derived across the cold front in order to confirm the cold front to be a contact discontinuity. We also derived the amount of suppression in the thermal conduction required to justify the existence of the cold front in A3667. We conclude by investigating the existence of a bow-shock which was previously suggested by \citet{vikhlinin01b}.

%--------------------------- Table 1 ------------------------------
\renewcommand{\tabcolsep}{3pt}
\begin{deluxetable*}{ccccc}
\tablewidth{0pt}
\tabletypesize{\small}
\tablecaption{Observation Details}
\tablehead{\colhead{ObsId} &  \colhead{Exposure (ks)} & \colhead{Clean Exposure (ks)} & \colhead{$z$} & \colhead{$N_H$ ($\times 10^{20}$)}} 
\tablecolumns{5}
\startdata
513,889,5751,5752   & 534 & 447 & 0.0556 & 4.71 \\ 
5753,6292,6295,6296 & ~   & ~   & ~   & ~    
\enddata
\tablecomments{Details of the archived Chandra data on Abell 3667 that has been used in this paper \citep{owers09b}.}
\label{tab:obs}
\end{deluxetable*}
%------------------------------------------------------------------

\section{X-ray data analysis}
\label{sec:anal}
We have used eight separate observations of Abell 3667 from the archival data of {\it Chandra} X-ray observatory. The details of the observations are listed in the Table~\ref{tab:obs}. We have developed a systematic calibration and analysis pipeline using {\it Chandra Interactive Analysis of Observations} (CIAO) and subsequent scripts in python and IDL. The details of the pipeline are discussed in \citet{schenck14}. For completeness, we outline some essential steps in the procedure here. 

\subsection{Calibration and Point Source Removal}

We have reprocessed the level 1 event files for all the eight Abell 3667 ACIS-I observations from the {\it Chandra} archive with CIAO version 4.3 and CALDB 4.4.6. These were the most updated versions available during the analysis. However, we have also tested more recent versions of CIAO and CALDB and found that the results are consistent. All the data for Abell 3667 considered for this paper were taken by ACIS-I detector. One of the observations (observation id: 889) is in the FAINT mode while the rest are in VFAINT mode. Backgrounds were taken into account using the blank-sky backgrounds in CALDB 4.4.6. The backgrounds were reprojected and processed to match the observations. Bad pixels and cosmic rays were removed using {\bf acis\_remove\_hotpix} and CTI corrections were made using {\bf acis\_process\_events}. Intervals of background flaring were excluded using light curves in the full band and 9-12 keV band. The light curves were binned at 259.3 seconds per bin, the binning used for the blank-sky backgrounds. Count rates greater than $3 \sigma$ from the mean were removed using {\bf deflare}. The light curves were visually inspected to ensure flares were effectively removed. 

Point sources were removed by first using the {\bf wavdetect} task in CIAO and then making manual corrections as necessary to ensure all contaminating point sources are excluded and no false detections were allowed. The regions which enclosed point sources were also removed from the backgrounds to prevent over-subtracting the background. The above steps were followed to create the 'CLEAN' data and background files for each observations separately.

\subsection{Combined X-ray Surface Brightness Map}
We have used the task {\bf merge\_all} in CIAO to combine all the CLEAN data files from eight observations and produce a combined surface brightness image. In order to reproject all the eight observations we have used the observation id : 5751 to be the reference. The surface brightness maps are produced after binning the CLEAN data by 4. The combined X-ray flux map is further adaptively smoothed (discussed in Section~\ref{sec:wvt}) and shown in Figure~\ref{fig:rgb}(a).  

\section{X-ray Temperature Maps}
\label{sec:temp}
The 'CLEAN' data from all eight observations where combined to produce a master count image. This master image is used to produce an 'outer-crop' region file which includes all the ACIS-I detectors from eight observations. This outer-crop region file is used to crop all the counts images (data and background) from individual observations. This is done in order to have all the counts images from individual observations cropped to the same dimensions and covering the same part of the sky. 

In order to make temperature maps from the combined data, we use two techniques: a) Weighted Voronoi Tessellation (WVT) method and b) Adaptive Circular Binning (ACB) method. In the following subsections, we will discuss in details the advantages and limitations of each of the two methods.

\subsection{WVT - Weighted Voronoi Tessellation Method}
\label{sec:wvt}
%------------------------ Figure:- 2 -----------------------------
\begin{figure*}[h!tb]
\centering
\epsfig{file=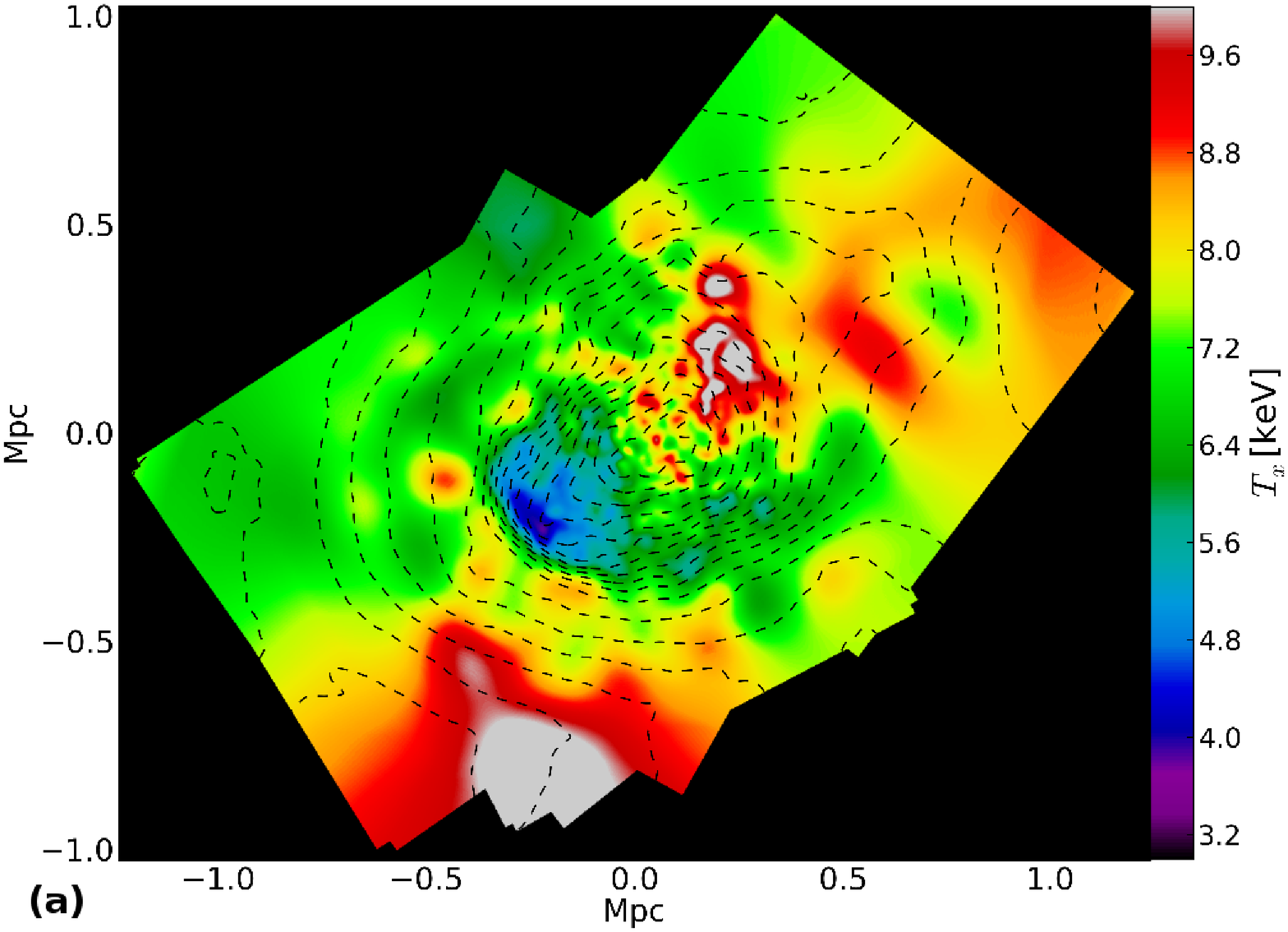,height=4.0truein}
\epsfig{file=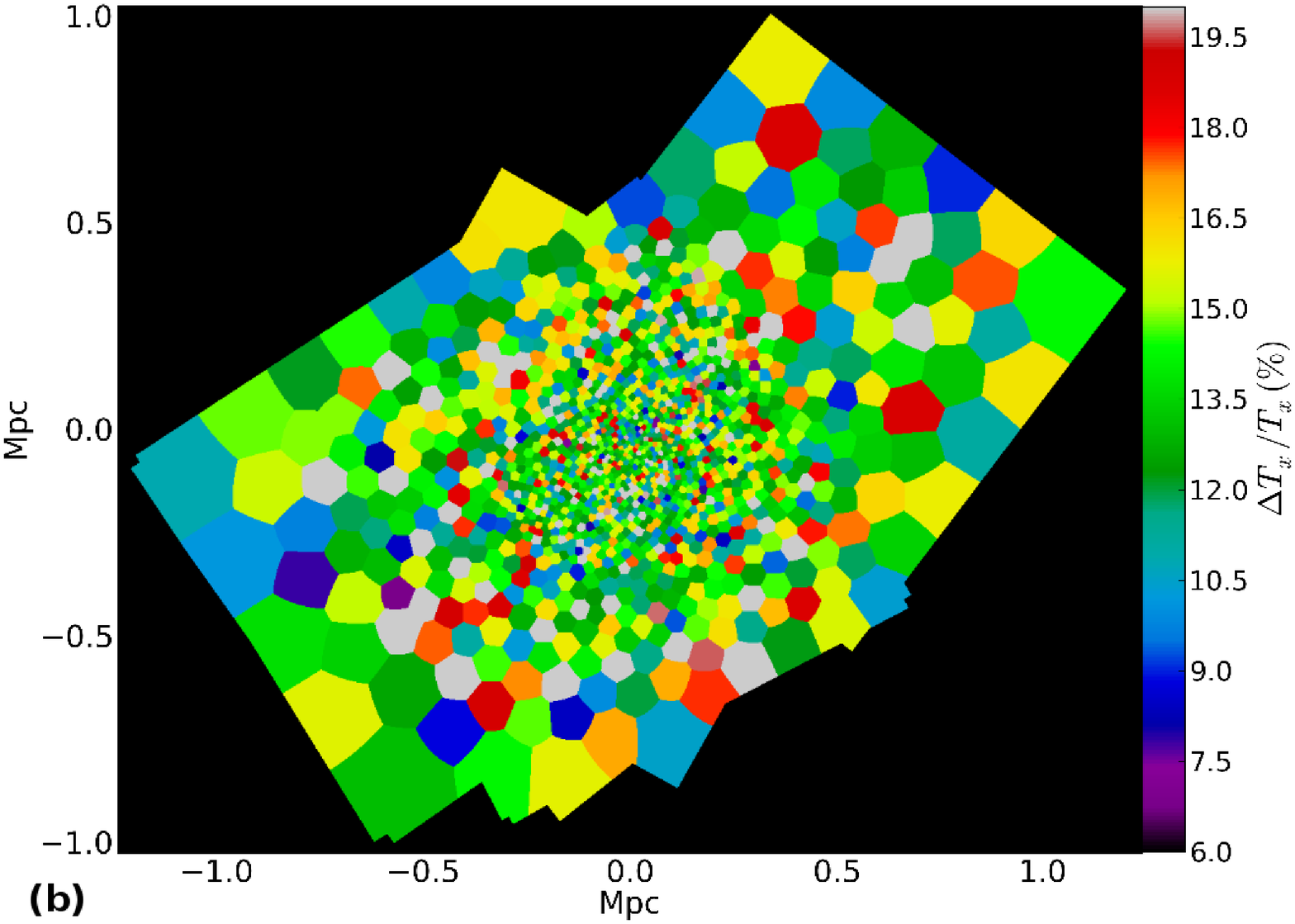,height=4.0truein}
\caption{{\bf (a)} Temperature map as derived from the WVT binning method and smoothed with the scalemap as obtained in ACB method (as in Section~\ref{sec:wvt}). Contours of the X-ray surface brightness map are overlayed on the temperature map. {\bf (b)} 1$\sigma$ errors for WVT temperature map expressed in percentage. The error map also shows the shape and location of the original WVT regions.}
\label{fig:Atemp_wvt}
\end{figure*}
%------------------------------------------------------------------

We use the Weighted Voronoi Tessellation (WVT) binning algorithm developed by \citet{diehl06}. The use of adaptive spatial binning using Voronoi Tessellations was first suggested by \citet{cappellari03} for Integral Field Spectroscopic data. Later, \citet{diehl06} generalized it for X-ray data analysis. WVT binning has been widely used in deriving X-ray temperatures \citep{pratt07,simionescu07,randall08,gastaldello09,rossetti10} and specifically to derive X-ray temperature maps for galaxy clusters [e.g. A4059 \citep{reynolds08}, A3128 \citep{werner07}, A2052 \citep{blanton09,deplaa10}, A2744 \citep{owers11}, A2254 \citep{girardi11}, A1201 \citep{ma12}, 1RXS J0603.3+4214 \citep{ogrean13}, CenA \citep{walker13}]. Our current work is the first attempt to derive an X-ray temperature map for A3667 using the WVT technique. 

Non-overlapping WVT regions are defined by the WVT-binning algorithm based on a given signal-to-noise ratio (SNR). The SNR is calculated after combining all eight CLEAN data and the CLEAN background files for A3667. The signal is equal to the background-subtracted counts and the noise assumes Poisson uncertainty contributions from both the source and background. The background counts have been rescaled by the ratio of the source and background exposure times for the WVT-binning operation. In this case, we have used a SNR of 50 to define the WVT regions. For the combined eight observations of A3667, WVT-binning produced $\sim 1400$ regions with a SNR scatter $\lesssim 4\%$.

Once the regions are defined, we use the {\it CIAO} task {\bf dmextract} to extract separate source and background spectra from the same region of the ACIS-I detector corresponding to each WVT region. For each WVT region, the source and background spectra are extracted for different observations separately. The weighted response matrix file (WRMF) and effective area function (WARF) are generated using the {\it CIAO} task {\bf specextract} using a region covering most of the field-of-view of the detector. A set of WRMF and WARF files are obtained for each observation. 

We have used {\it XSPEC} 12.8 to perform spectral fitting between 0.7 to 8.0 keV. The {\it APEC} thermal plasma model along with the {\it PHABS} photo-electric absorption model were used to fit the spectra from each WVT region. The redshift of the cluster was kept fixed at $z=0.055$. The hydrogen column density was fixed at $N_H=4.7\times 10^{18}~cm^{-2}$, as obtained from the LAB survey \citep{kalberla05} near the location of the cluster. In order to combine all eight observations, we have simultaneously fit all eight spectra from each WVT region in {\it XSPEC} using the C-statistic \citep{cash79}. 

The metallicity of the cluster was kept fixed at $0.3Z_{\odot}$ \citep{lovisari09}. Only the temperature and the {\it APEC} normalization were fit for each region. The WVT temperature map along with the error map were obtained. Later, the temperature map was adaptively smoothed using a variant of the {\it ASMOOTH} program developed by \citet{ebeling06}. The adaptively smoothed WVT temperature map is shown in Figure~\ref{fig:Atemp_wvt}(a). Figure~\ref{fig:Atemp_wvt}(b) shows the average percentage errors ($\Delta T_x/T_x \times 100$) in the best-fitted temperature values ($T_x$). This error map also shows the exact location and size of the WVT regions. The errors are estimated inside {\it XSPEC} using a Markov Chain Monte Carlo technique with chain length of 10,000. The plus ($\sigma_{+}$) and minus ($\sigma_{-}$) errors are calculated at $1\sigma$ level. The resultant error $\Delta T_x$ is the mean of $\sigma_{+}$ and $\sigma_{-}$. It should be noted that $\gtrsim 90\%$ of the regions have $\lesssim 20\%$ errors in temperature. Figure~\ref{fig:Aterr_dist} shows the distribution of the $\Delta T_x/T_x$ with respect to the best-fitted $T_x$ values. From this figure we can conclude that there is a definite trend confirming that the regions with higher temperatures have higher errors. This is expected as the {\it Chandra} instrument response deteriorates at higher energies. Figure~\ref{fig:Aterr_dist} also shows that the distribution of $\sigma_{-}$ and $\sigma_{+}$ are different. 

We compared our WVT temperature map with the existing X-ray temperature maps by \citet{briel04}, \citet{lovisari09} and \citet{finoguenov10} from XMM Newton observations. Our adaptively smoothed WVT temperature map (Figure~\ref{fig:Atemp_wvt}(a)) recovers the basic features seen in previous XMM studies \citep{briel04,lovisari09}. However, the higher exposure time for the combined {\it Chandra} observations allowed us to derive a higher resolution temperature map than the above studies. The A3667 temperature map from \citet{owers09b} also combined all the eight {\it Chandra} observations, but their temperature map is limited to a very small region around the cold front. The temperature values near the cold front in our current WVT map (Figure~\ref{fig:Atemp_wvt}(a)) are in agreement with that in \citet{owers09b}. The lowest temperature in Figure~\ref{fig:Atemp_wvt}(a) occurs near the inner edge of the cold front and is about $4.0$~keV. There is a high temperature region behind the cold front towards the North-West where temperatures rise up to $\sim 11$~keV. 

The WVT method creates non-overlapping regions within which a spectrum is extracted. So the temperatures between adjacent WVT regions are independent. This is advantageous in propagating errors in fitted temperatures to derived quantities like temperature jumps, Mach number, pressure, entropy, etc. The major limitation of the WVT method is that temperature structure depends on the WVT bin locations. These bin locations are not unique and depend on the initial conditions of the bin accretion in the WVT-binning algorithm \citep{diehl06}. Hence, a slight change in the input SNR criteria or the starting location of the bin-accretion can create slightly different WVT temperature maps. Since the bins are non-overlapping, the WVT method might miss a feature that is lower than the size of the bin at particular location on the map.  In order to verify whether we have missed any crucial feature in the A3667 WVT temperature due to this limitation of the algorithm, we have used another technique that we call Adaptive Circular Binning (ACB) in the next section.

\subsection{ACB - Adaptive Circular Binning Method :}
\label{sec:acb}
%------------------------ Figure:- 3 -----------------------------
\begin{figure*}[h!tb]
\centering
\epsfig{file=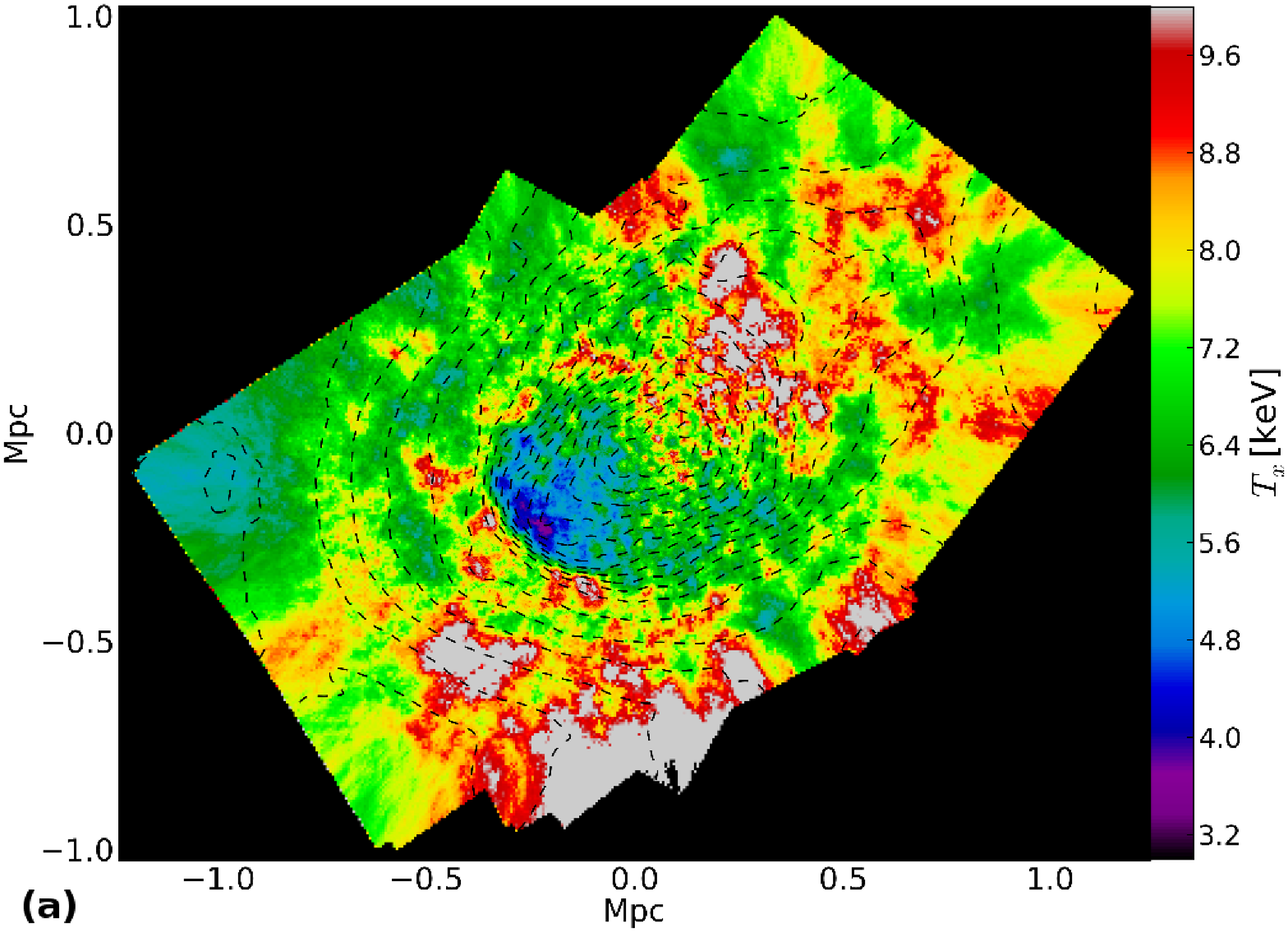,height=4.0truein}
\epsfig{file=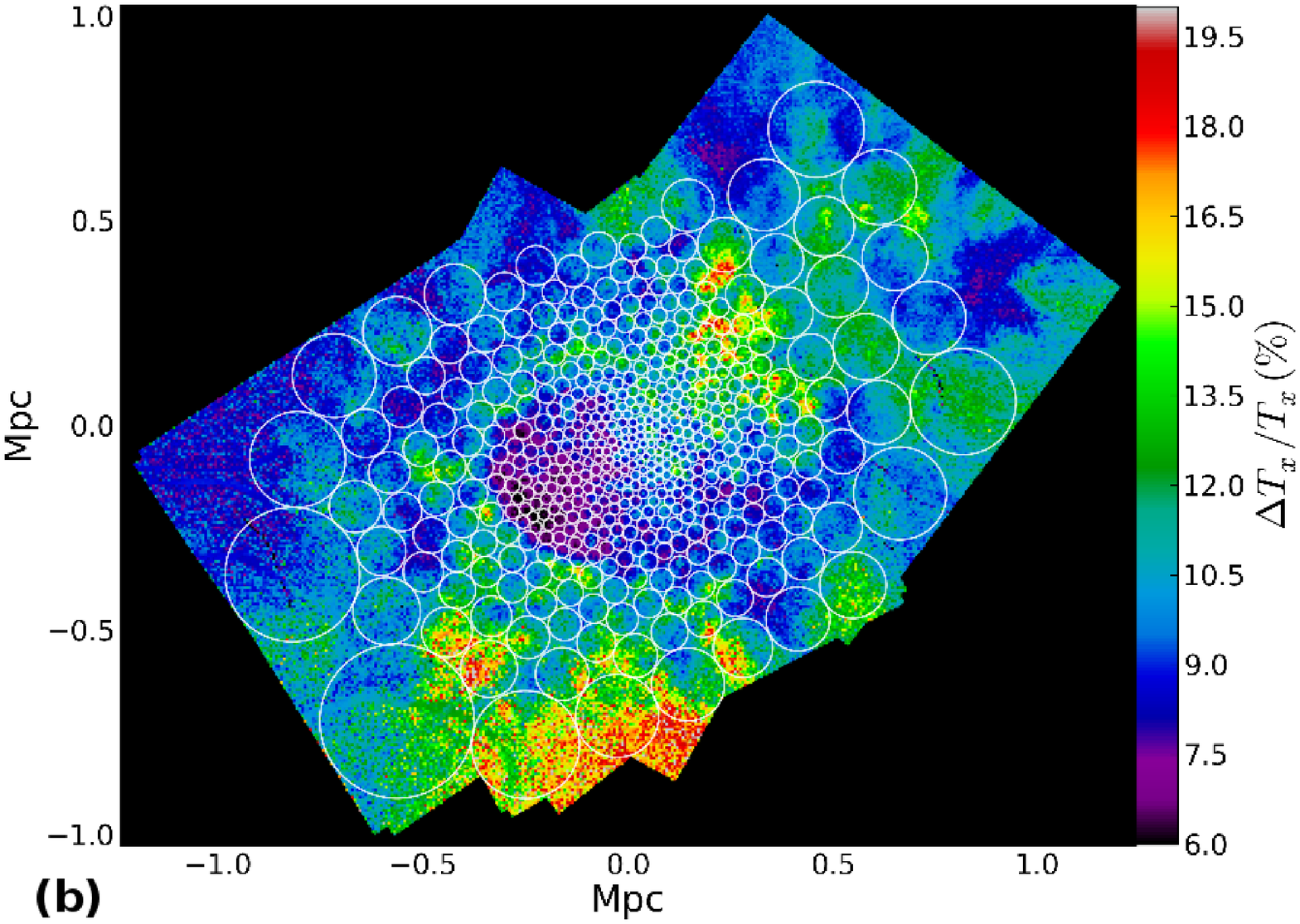,height=4.0truein}
\caption{{\bf (a)} Temperature map as derived from the ACB method (as in Section~\ref{sec:acb}). Contours of the smoothed X-ray surface brightness map is overlayed on both the images. {\bf (b)} Corresponding $1\sigma$ error map expressed in percentages. Some of the ACB circles are overlayed on the error map to show the characteristic scales of binning in each location of the A3667 map.}
\label{fig:Atemp_acb}
\end{figure*}
%------------------------------------------------------------------
%------------------------ Figure:- 4 -----------------------------
\begin{figure}[h!tb]
\centering
\epsfig{file=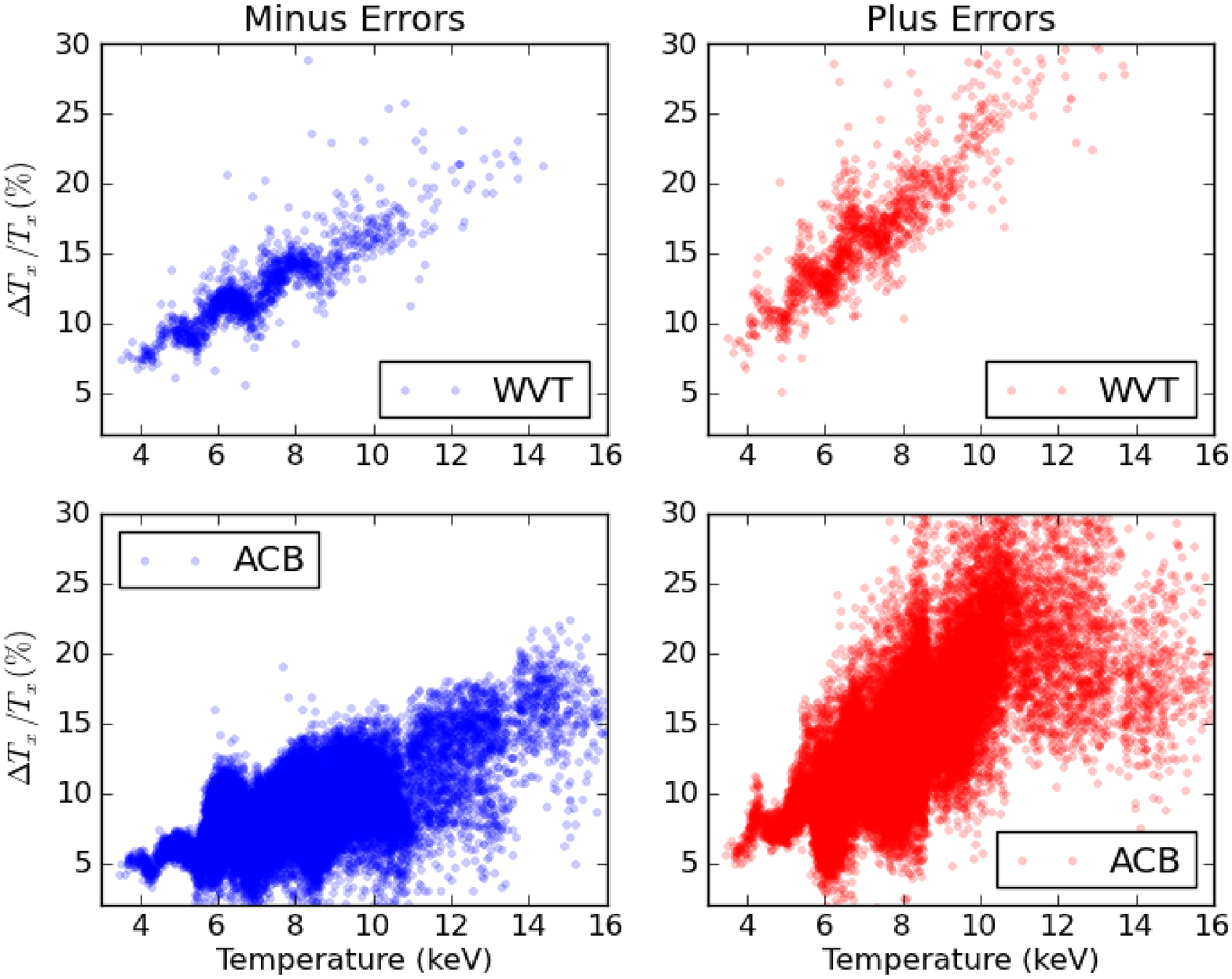,height=2.6truein}
\caption{{\bf Top row:} The distribution of the $\sigma_{-}$ and $\sigma_{+}$ errors with respect to the best-fitted $T_x$ values as obtained for the WVT temperature map (Figure~\ref{fig:Atemp_wvt}). The error estimation procedure is detailed in section~\ref{sec:wvt}. {\bf Bottom row:} The distribution of the $\sigma_{-}$ and $\sigma_{+}$ errors with respect to the best-fitted $T_x$ values as obtained for the ACB temperature map (Figure~\ref{fig:Atemp_acb}).}
\label{fig:Aterr_dist}
\end{figure}
%------------------------------------------------------------------

The ACB method is based on extracting spectra from a circular region centered on a particular pixel in the map. Unlike the WVT method, the ACB circles from adjacent pixels can significantly overlap with each other. The size of the ACB circles depends on the input SNR criteria (similar to the WVT regions). It should be noted that X-ray temperature maps have been previously produced using a technique similar to the ACB method [e.g. M86 \citep{randall08}, A133 \citep{randall10}, A2744 \citep{owers11}, A2443 \citep{clarke13}]. Previous works have used either a cut-off in counts or background subtracted counts to derive the circular region. In order to be consistent with the WVT method, we derive ACB circles based on the SNR criteria (same as in WVT method). 

The steps followed in the ACB method include:
\begin{enumerate}
\item Use the native pixels in the combined counts images for both the CLEAN source and background images.
\item For each pixel, we define a circular region (centered on that pixel). The radius of a circle is such that the SNR calculated from the source and background images is about 50 (same as in WVT binning method). The signal is equal to the background-subtracted counts and the noise assumes Poisson uncertainty contributions from both the source and background.
\item Spectra from both the CLEAN source and background data (from each observation) are then extracted for the corresponding circular region for that pixel. The spectral extraction is done with {\it CIAO} task {\bf dmextract}. 
\item The weighted response matrix file (WRMF) and effective area function (WARF) are
generated using the {\it CIAO} task {\bf specextract} using a region covering most of the field-of-view of the 
detector. A set of WRMF and WARF files are obtained for each observations.
\item For each pixel, we then simultaneously fit spectra for all eight observations using {\it XSPEC} 12.8. 
\item The {\it APEC} thermal plasma model along with the {\it PHABS} photo-electric absorption model were used to fit the spectra from each pixel between 0.7 and 8.0 keV. 
\item The redshift of A3667, the $N_H$ value and the metallicity are kept fixed for {\it XSPEC} fitting at the same value as in WVT binning method. The temperature and the {\it APEC} normalization were kept free to be fitted (same as in section~\ref{sec:wvt}).
\end{enumerate} 

Apart from the process of defining the ACB regions, most of the other steps are common between both ACB and WVT. In the case of ACB, the total number of regions are equal to the total number of image pixels within the combined detector area. For A3667, ACB method produced $\sim 5.9\times10^5$ regions. All the initial data analysis within {\it CIAO}, to produce the CLEAN data-sets, as well as the WVT temperature maps were carried out in a 32 core cluster machine with a total of 64 Gb shared RAM. To process these large number of ACB regions, we used $\gtrsim 1600$~cores of the JANUS supercomputer at University of Colorado Boulder for a total of $\sim 4-6$~hours.  

%------------------------ Figure:- 5 -----------------------------
\begin{figure*}[t!]
\centering
\epsfig{file=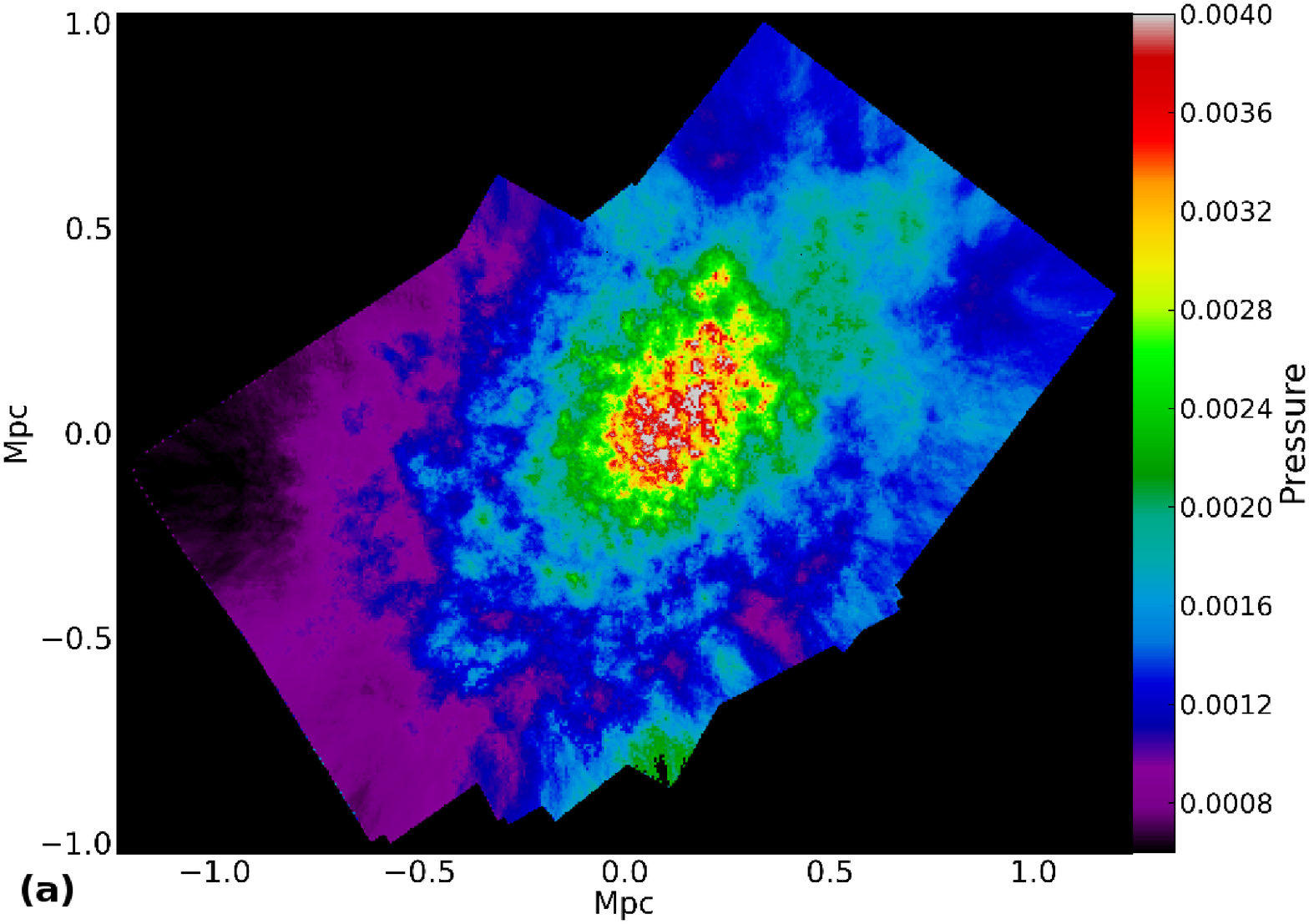,height=4.2truein}
\epsfig{file=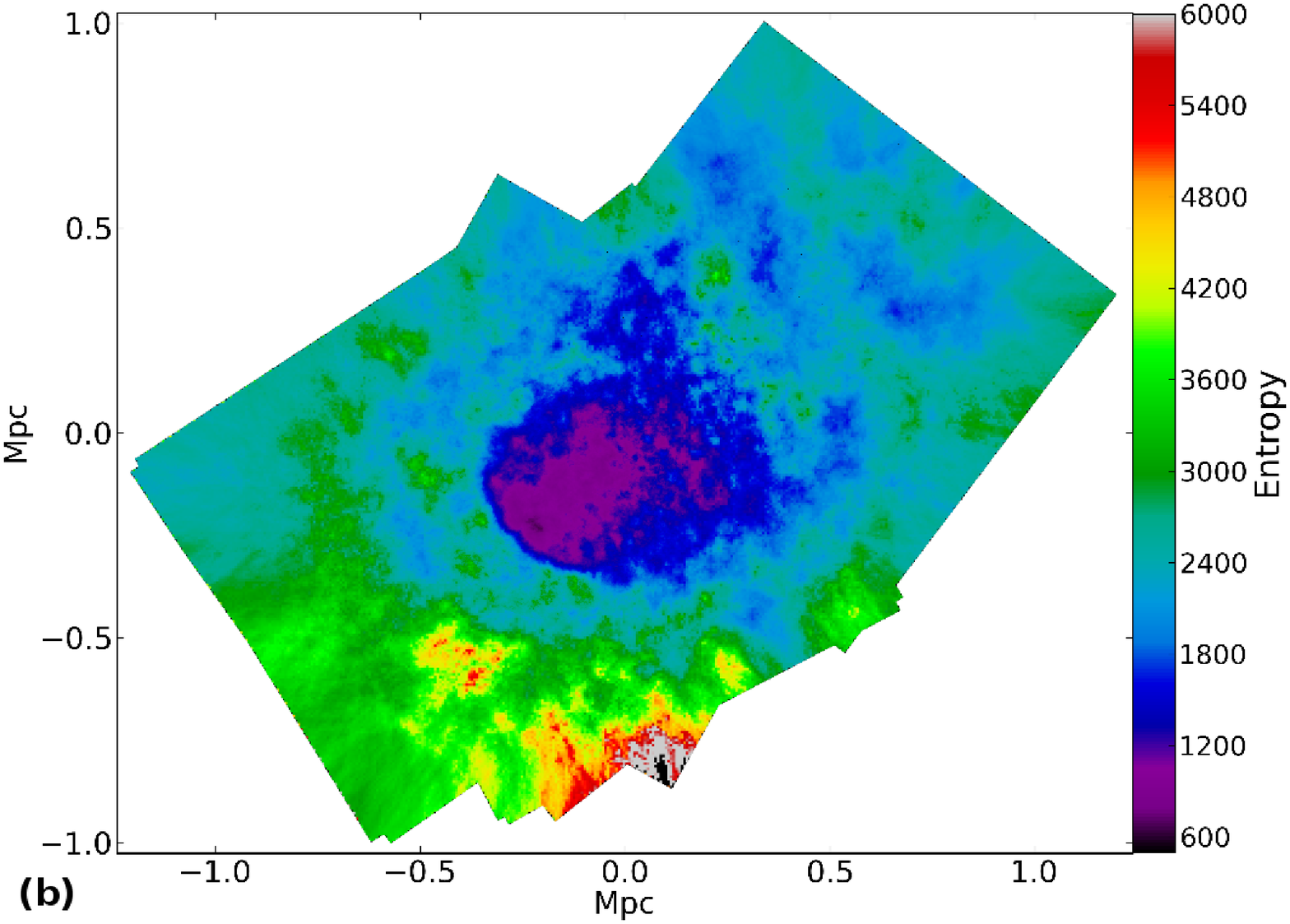,height=4.2truein}
\caption{{\bf (a)} Pseudo-pressure map evaluated as $P=\sqrt{S_x}T_x$. {\bf (b)} Pseudo-entropy map evaluated as $K_s=T_x/S_x^{1/3}$. For both the map $S_x$ is same as in Figure~\ref{fig:rgb}(a) and  $T_x$ is from ACB method as in Figure~\ref{fig:Atemp_acb}(a).}
\label{fig:APE}
\end{figure*}
%------------------------------------------------------------------

While WVT produces regions with no overlap, the ACB region for a pixel has significant overlap with the ACB regions of the nearby pixels. The overlap area depends on the SNR around that pixel location on the detector. The ACB temperature map is shown in Figure~\ref{fig:Atemp_acb}(a). This temperature map is not smoothed further unlike the WVT temperature map. The corresponding $1\sigma$ errors ($\Delta T_x/T_x \times 100$), in percentages, are shown in Figure~\ref{fig:Atemp_acb}(b). It should be noted that $\gtrsim 90\%$ of the pixels have $\lesssim 20\%$ errors in temperature. The errors are estimated in the same manner as in section~\ref{sec:wvt}. Figure~\ref{fig:Aterr_dist} shows the distribution of the $\sigma_{-}$ and $\sigma_{+}$ errors with respect to the best-fitted $T_x$ values. This again confirms that pixels with higher temperatures have higher errors (same as in section~\ref{sec:wvt}). The ACB temperature map has the overall same features as the WVT image (Figure~\ref{fig:Atemp_wvt}(a)). In Figure~\ref{fig:Atemp_acb}(b) we have overlayed some of the ACB circles on the error map. It should be noted that the scales of the ACB circles are minimum near the inner edge of the cold front and gradually grow to the outskirts of the field-of-view. Comparing Figures~\ref{fig:Atemp_wvt} and \ref{fig:Atemp_acb}, we can confirm that the WVT regions are small enough to be able capture the overall temperature structure of the A3667 like the ACB method. This is possible due to the high SNR for the combined observations of A3667. The temperature values in Figure~\ref{fig:Atemp_acb}(a) are in agreement with the X-ray temperature map of A3667 in \citet{owers09b}, where they used a method similar to ACB. However, our current ACB temperature map has a higher resolution near the cold front and larger field-of-view than that in \citet{owers09b}. We have extracted spectra for every pixel of the map around the cold front in order to get the sharpest profile of thermodynamic quantities across the cold front. It is possible that \citet{owers09b} might have used a lower resolution map to extract fewer number of spectra. Comparing our current ACB map with \citet{owers09b} as well as previous XMM studies \citep{briel04,lovisari09, finoguenov10}, we conclude that our ACB temperature map has a combination of the highest resolution and largest field-of-view produced to date for A3667. The magnitude of the error map also supports this map to be the highest fidelity X-ray temperature map for A3667. Unfortunately, {\it Chandra} observations do not extend to the North-West relic region. Hence, XMM observations by \citet{finoguenov10} is the only study to estimate X-ray temperatures and shocks near the radio relic.

In Figure~\ref{fig:APE} we show the pseudo-pressure and pseudo-entropy maps derived from the smoothed X-ray surface brightness map and the ACB temperature map. The pseudo-pressure is calculated as $P=\sqrt{S_x}T_x$ and pseudo-entropy is calculated as $K_s=T_x/S_x^{1/3}$. The pseudo-pressure shows a very smooth variation near the cold front, confirming it as a contact discontinuity. This will be more clear in Section~\ref{sec:cf} when we will derive the actual pressure profile across the cold front. On the other hand, it is clear from the pseudo-entropy map that there is a strong jump near the location of the cold front. Entropy is lower inside the cold front as compared to the outside. The pseudo-entropy map also confirms the cold front. The resolution of the entropy map is much higher than seen in previous studies with XMM-Newton \citep{briel04}.

\section{Shock Finder}
\label{sec:sf}

%------------------------ Figure:- 6 -----------------------------
\begin{figure*}[t!]
\centering
\epsfig{file=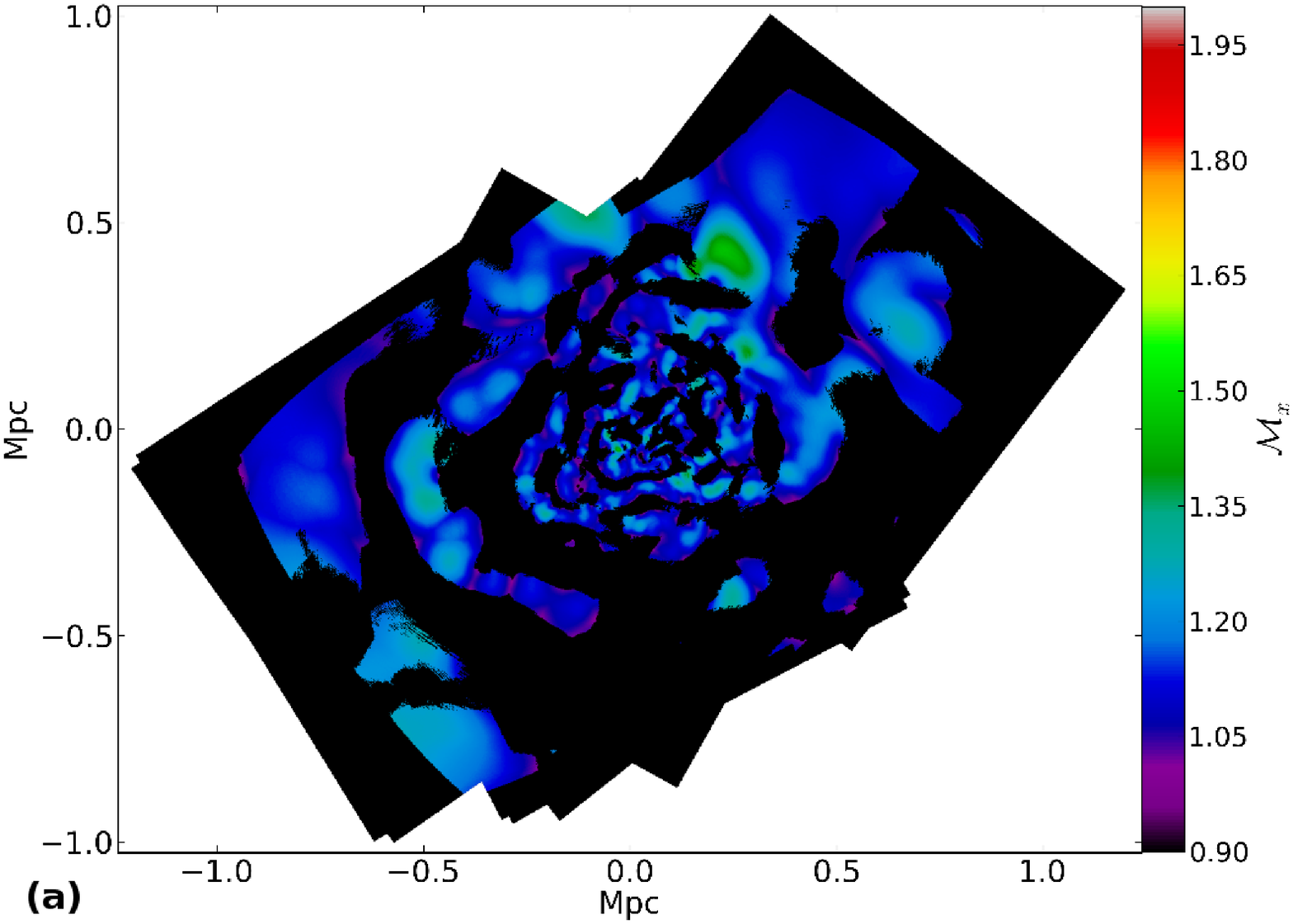,height=4.2truein}
\epsfig{file=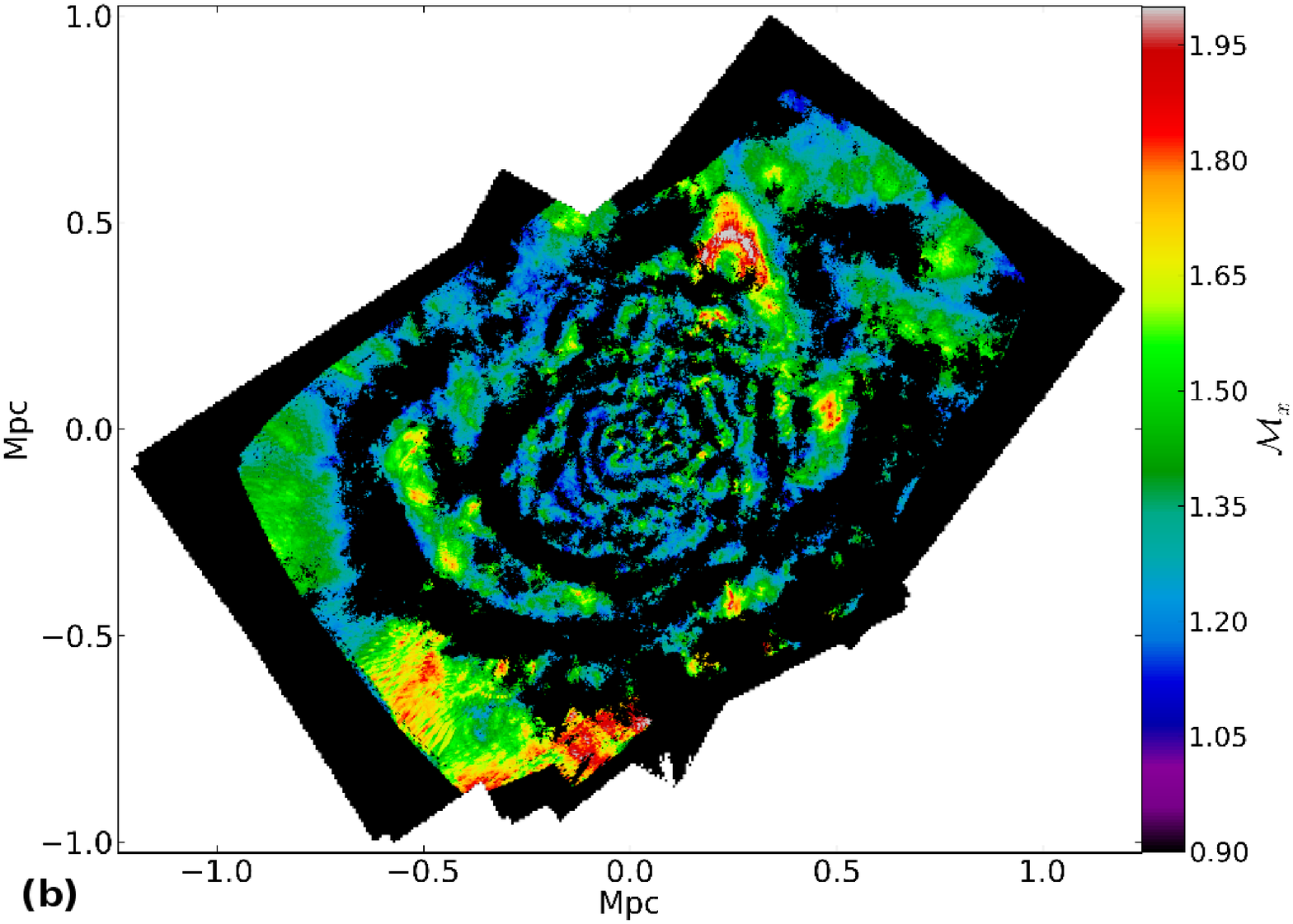,height=4.2truein}
\caption{{\bf (a)} Mach number map as obtained by applying the shock-finder on the WVT temperature map [as in Figure~\ref{fig:Atemp_wvt}(a)] and the surface brightness map (Figure~\ref{fig:rgb}(a)). {\bf (b)} Mach number map as obtained by applying the shock-finder on the ACB temperature map [as in Figure~\ref{fig:Atemp_acb}(a)] and the surface brightness map (Figure~\ref{fig:rgb}(a)). } 
\label{fig:Amach}
\end{figure*}
%------------------------------------------------------------------
%------------------------ Figure:- 7 -----------------------------
\begin{figure}[t!]
\centering
\epsfig{file=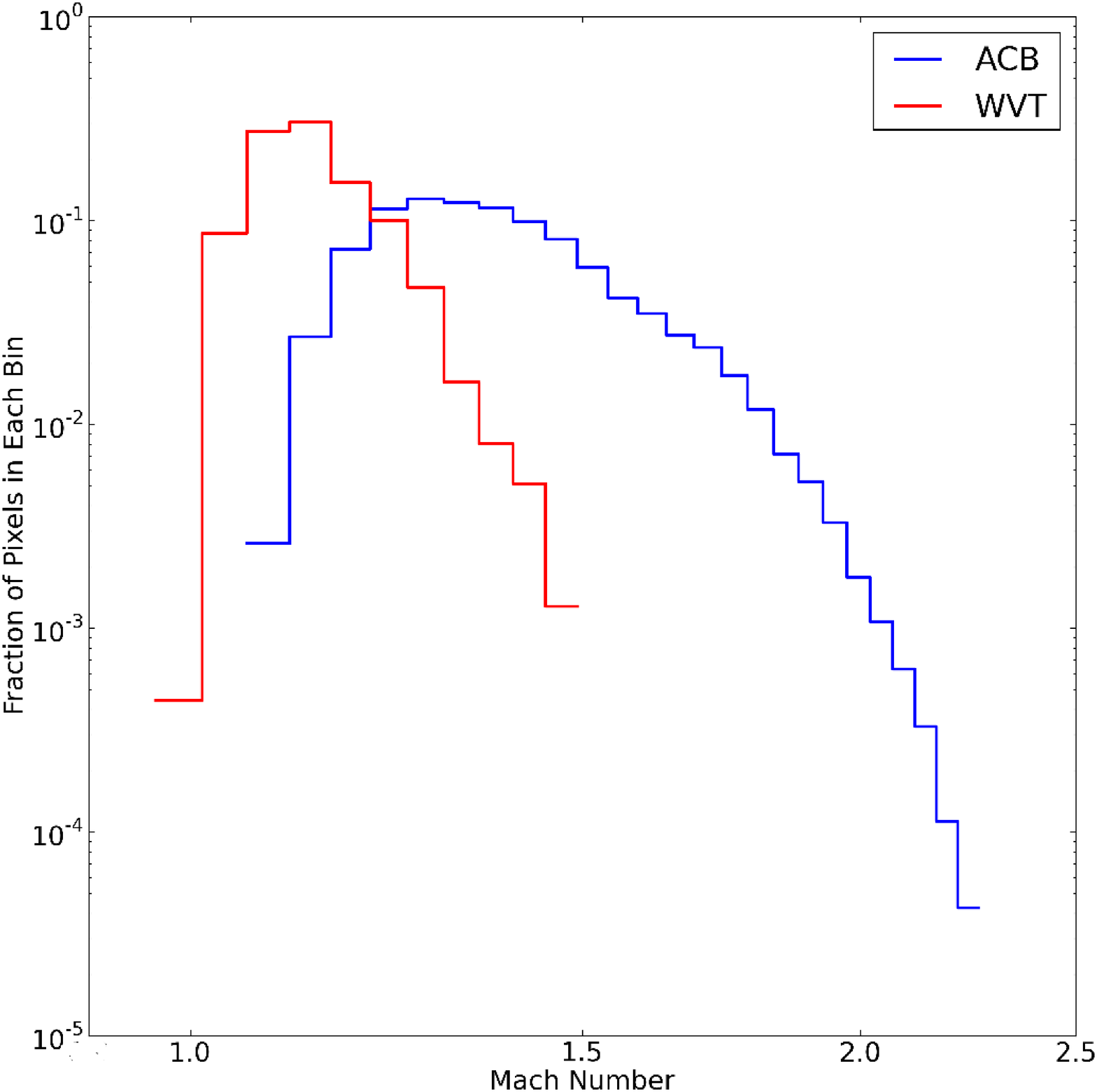,height=3.2truein}
\caption{Mach number distribution from X-ray Mach number maps (Figure~\ref{fig:Amach}) for both the WVT and ACB methods.}
\label{fig:Ahist}
\end{figure}
%------------------------------------------------------------------
Galaxy cluster formation is a violent process in which subclusters crash onto clusters at generally supersonic velocities creating shocks. Shocks are fundamental in understanding the baryonic processes in the clusters because they convert gravitational potential energy into thermal energy, thus heating the ICM \citep{roettiger96,burns98} and allowing the cluster gas to approach hydrostatic equilibrium. There are mainly two types of shocks in the hot ICM plasma. Accretion shocks arise at larger distances ($\gtrsim$ several Mpc) from the cluster center. Cosmological simulations \citep{skillman08,vazza09} predict that the much cooler IGM (Inter-galactic Medium) accretes onto the clusters through a system of shocks producing Mach numbers $\mathcal{M} \sim 10-100$ \citep{miniati00,ryu03}. The merger shocks arise as sub-halos fall into the main clusters producing moderate shocks with $\mathcal{M} \lesssim 5$ \citep{paul11}.

One of the major goals of this paper is to study the X-ray shocks using the highest resolution temperature maps derived in the previous section. In order to systematically find shocks across the entire field-of-view of the combined {\it Chandra} observations we developed an automated shock-finder. This is adapted from the method described in \citet{skillman08,skillman13} for numerical simulations of galaxy clusters where they use a systematic and unbiased method of searching for shocks in the 3-dimensional cluster environment. Their technique relies on a temperature-jump based shock-finder as described in \citet{skillman08}. The temperature jump is preferable to density jump as the former is more sensitive to the Mach number. In contrast, the density jump quickly asymptotes for strong shocks \citep{skillman08}.

The Rankine-Hugoniot temperature jump condition is used to calculate the Mach number as:
\begin{equation}\label{eq:tempjump}
\frac{T_2}{T_1}=\frac{(5\M^2 - 1)(\M^2+3)}{16\M^2}.
\end{equation}
In the 3-dimensional shock-finding algorithm, used in cosmological simulations, a pixel is determined to have a shock if it meets the following criteria \citep{skillman08}:
\begin{eqnarray}\label{eq:jump_cond}
\nabla \cdot {\bf v} & < & 0 \nonumber \\
\nabla{T}  \cdot \nabla{K_S} & > & 0 \\
 T_2 &>& T_1 \nonumber \\
\rho_2 & > & \rho_1 \nonumber
\end{eqnarray}
where ${\bf v}$ is the velocity field, $T$ is the temperature, $\rho$ is the density and $K_S=T/\rho^{\gamma -1}$ is the entropy. At each pixel, the 3-dimensional shock-finder calculates the jump in temperature and surface brightness in both grid directions. A pixel is marked as a shock if the temperature gradient and surface brightness gradient have the same sign since the temperature and density both increase from pre-shock to post-shock. The Mach number is then calculated from the temperature jump using equation~\ref{eq:tempjump}. This 3-dimensional shock-finder is applied to a simulated MHD cluster (discussed in Section~\ref{sec:sim}). 

Since the observations are restricted to the plane of the sky, we have modified the 3-dimensional shock-finder to work on 2-dimensional projected X-ray surface brightness and temperature maps. In this modified scenario, the shock-finder calculates the jump in temperature and surface brightness in $N$ evenly placed directions centering on a given pixel. For A3667, we have used $N=64$. The shock-finder then accepts those pixel-pairs between which the conditions $T_2>T_1$ and $S_{X1} > S_{X1}$ (a proxy for $\rho_2 > \rho_1$ in observations) in equation~\ref{eq:jump_cond} are satisfied. The other two conditions, $\nabla \cdot {\bf v} < 0$ and $\nabla{T}  \cdot \nabla{K_S} > 0$, in equation \ref{eq:jump_cond} cannot be used in the case of observations. The Mach number for each successful pixel-pair is noted. The Mach number with the maximum value is chosen to be the resultant Mach number for that given pixel. To choose the pixel-pairs in each direction we use a 'jump-length' from the corresponding scalemap, which gives the size of the Voronoi region or the radius of the ACB circle for that given pixel. A fiducial multiplicative factor is also applied on the scale-lengths to define the 'jump-length'. The multiplicative factor has been varied to assess the scale appropriate for detecting the shocks. In these cases, we found the factor of 1.25 is appropriate for the Mach number estimation. Detailed description of this observational shock-finder is discussed in \citet{schenck14}.

The resultant Mach number map for A3667 is shown in Figure~\ref{fig:Amach} for both the WVT and ACB temperature maps. The Mach numbers are $\lesssim 2$ in all cases and consistent with those found in the cores of numerically simulated clusters \citep{skillman08}. For our observational shock-finder, the resolution of the Mach number map directly depends on the resolution of the temperature map. Since the ACB map has a higher resolution, the resolution of the Mach number map for ACB is higher than that in the case of WVT.  Moreover, the ACB Mach numbers are mostly around 1.5-1.6, whereas that in WVT are mostly between 1.1-1.2. The comparison between the Mach number distributions from WVT and ACB temperature maps are shown in Figure~\ref{fig:Ahist}. Since the Mach number distribution from the WVT method suffers strongly from the size scales of the WVT bins and are mostly restricted below Mach number of 1.5, we will use the Mach number distribution from the ACB temperature map for future comparison with the simulations in section~\ref{sec:compXRS} 

In order to validate the observational shock-finder, we will compare it with the original shock-finder which operates on the 3-dimensional cosmological simulations in Section~\ref{sec:compXRS}. Since the combined {\it Chandra} observations for A3667 do not extend beyond the central $1$~Mpc region around the center of the cluster, we will use previous radio observations of A3667 \citep{huub97} which extend to the outskirts of the cluster. In the following sections, we will compare the combined shock statistics from X-ray and radio data of A3667 with the results from the 3-dimensional shock-finder on the cosmological AMR MHD simulations.

\section{Simulations}
\label{sec:sim}
%------------------------ Figure:- 8 -----------------------------
\begin{figure*}[t!]
\centering
\epsfig{file=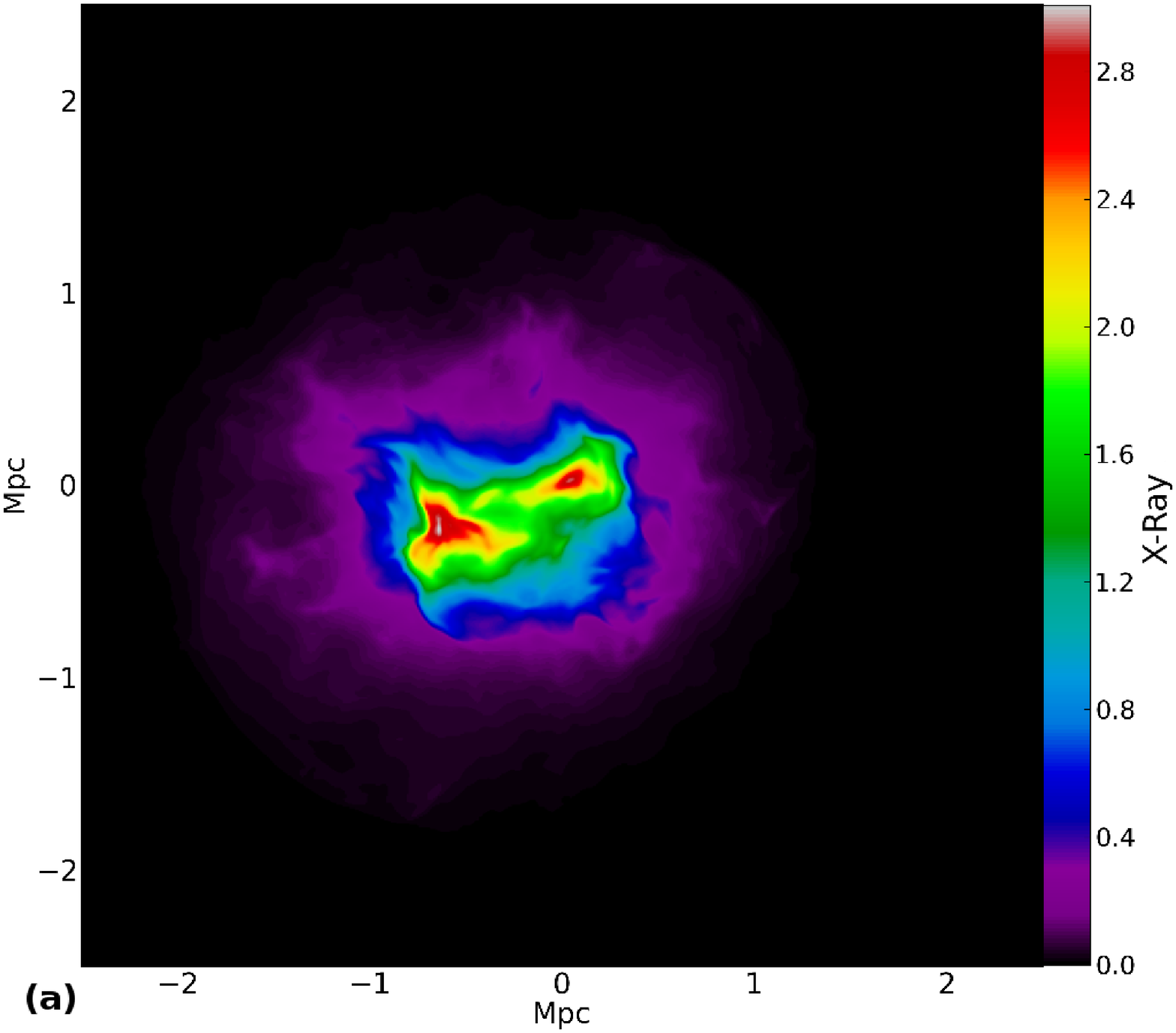,height=3.0truein}
\epsfig{file=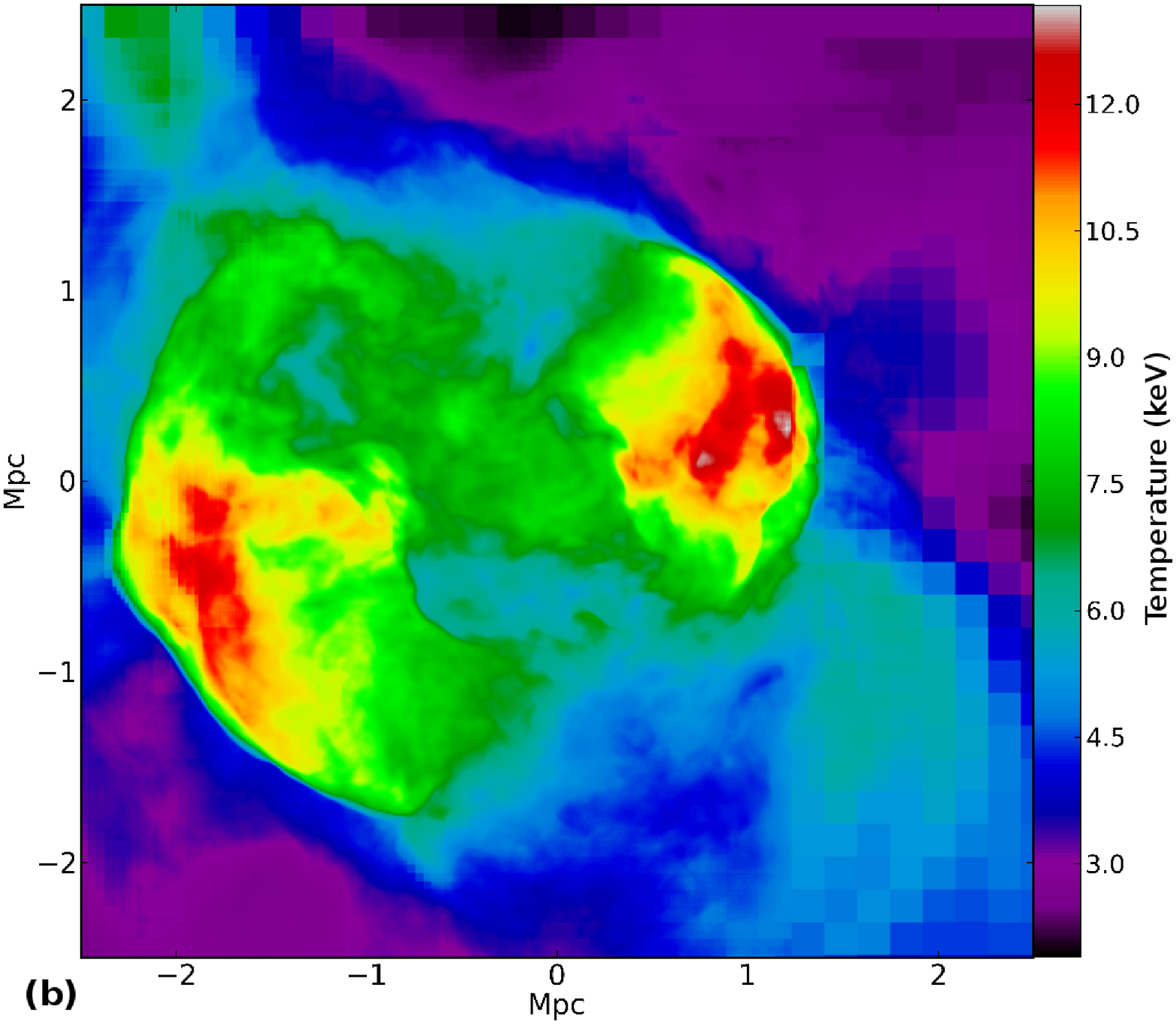,height=3.0truein}
\caption{{\bf (a)} Projected X-ray surface brightness map as obtained from the MHD simulations. {\bf (b)} Projected X-ray temperature map as obtained from the MHD simulations [see \citet{skillman13}].}
\label{fig:simXT}
\end{figure*}
%------------------------------------------------------------------
%------------------------ Figure:- 9 -----------------------------
\begin{figure*}[t!]
\centering
\epsfig{file=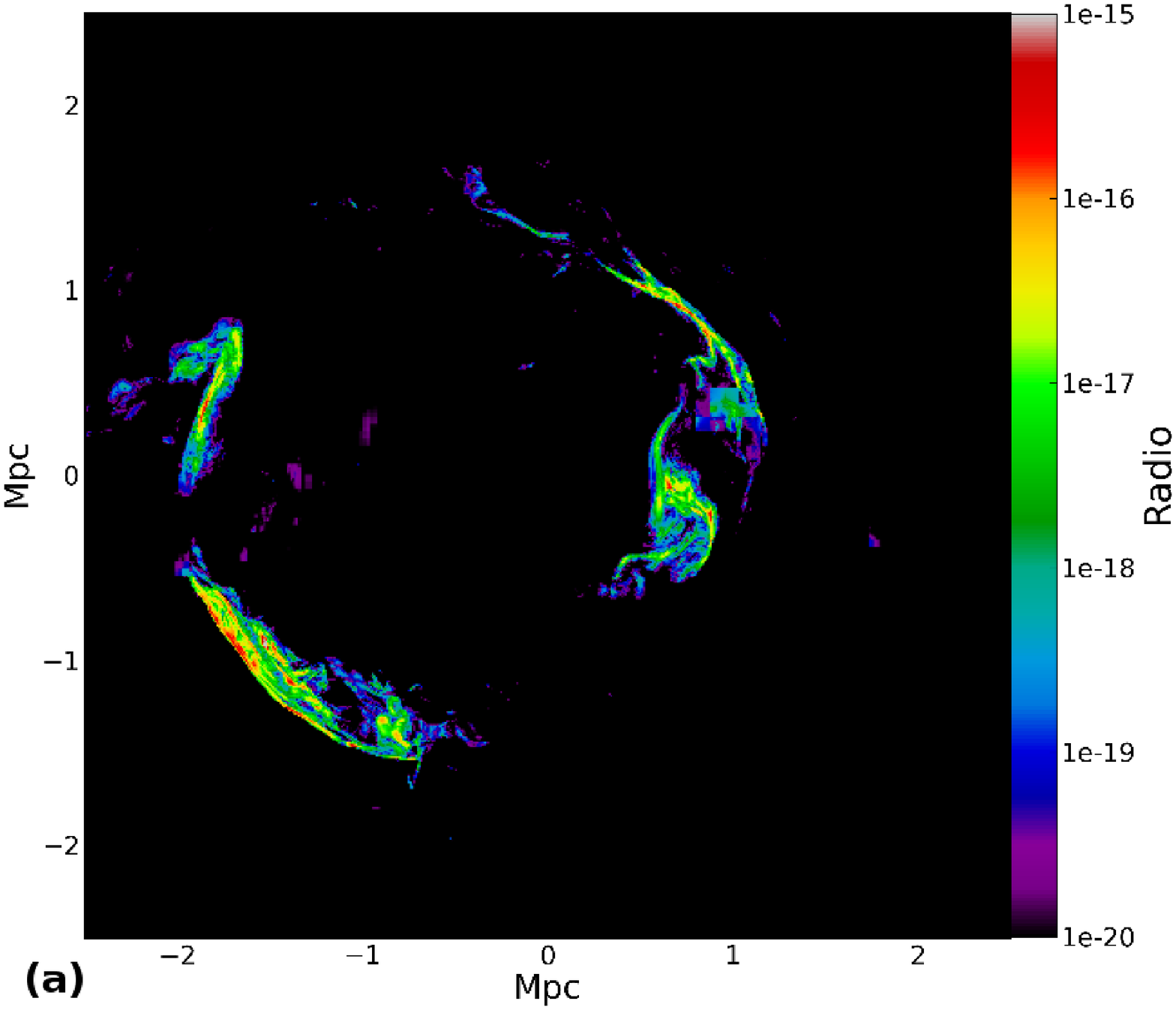,height=3.0truein}
\epsfig{file=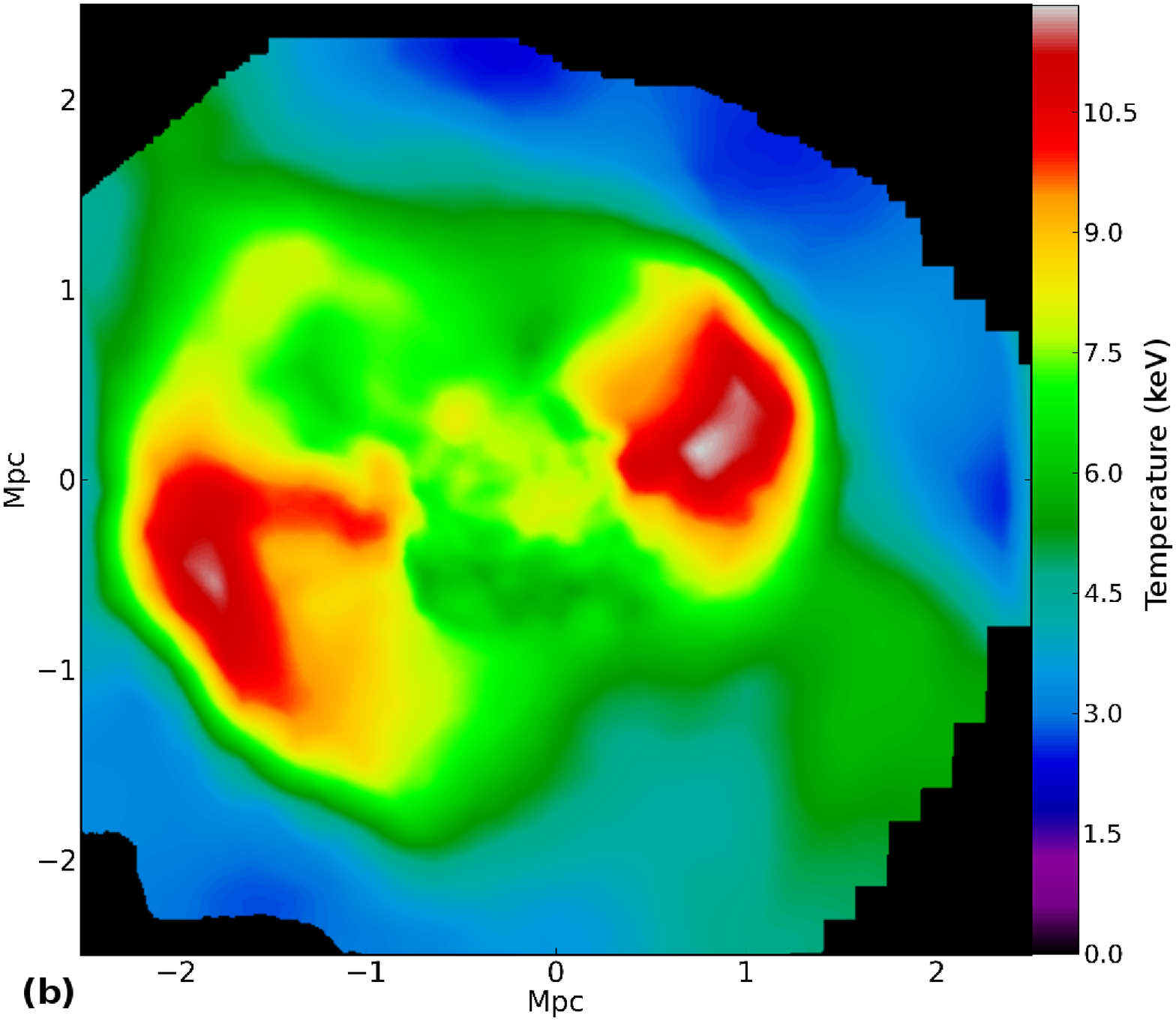,height=3.0truein}
\caption{{\bf (a)} Radio emission at 1.4 GHz as obtained from the MHD simulations \citep{skillman13}. {\bf (b)} Projected, WVT-binned, smoothed X-ray temperature map as obtained from the MHD simulations. The projected temperature map has been binned and smoothed based on the WVT kernel similar to that used for A3667 as described in Section~\ref{sec:wvt}.}
\label{fig:simXT_wvt}
\end{figure*}
%------------------------------------------------------------------
%------------------------ Figure:- 10 -----------------------------
\begin{figure*}[t!]
\centering
\epsfig{file=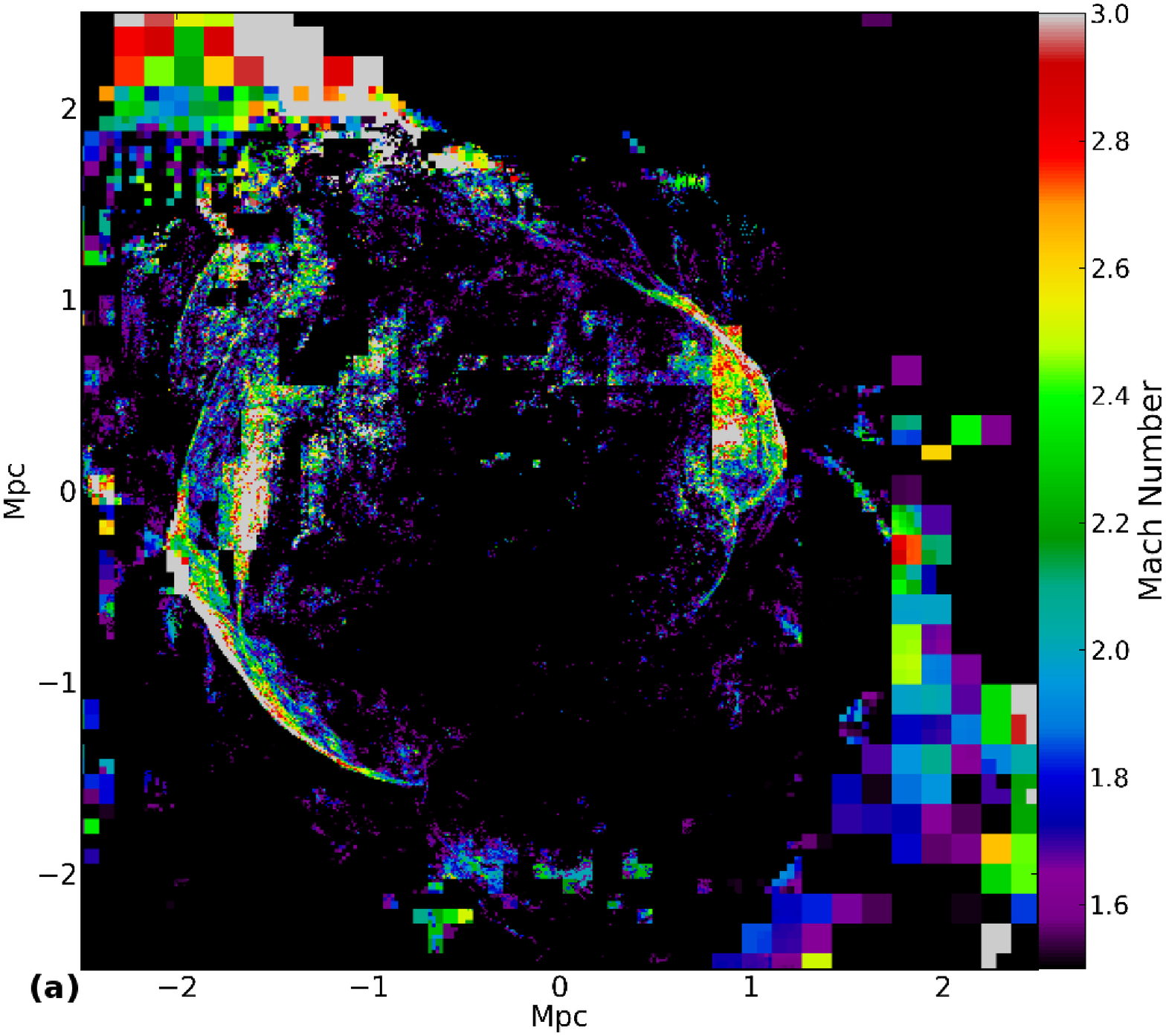,height=3.0truein}
\epsfig{file=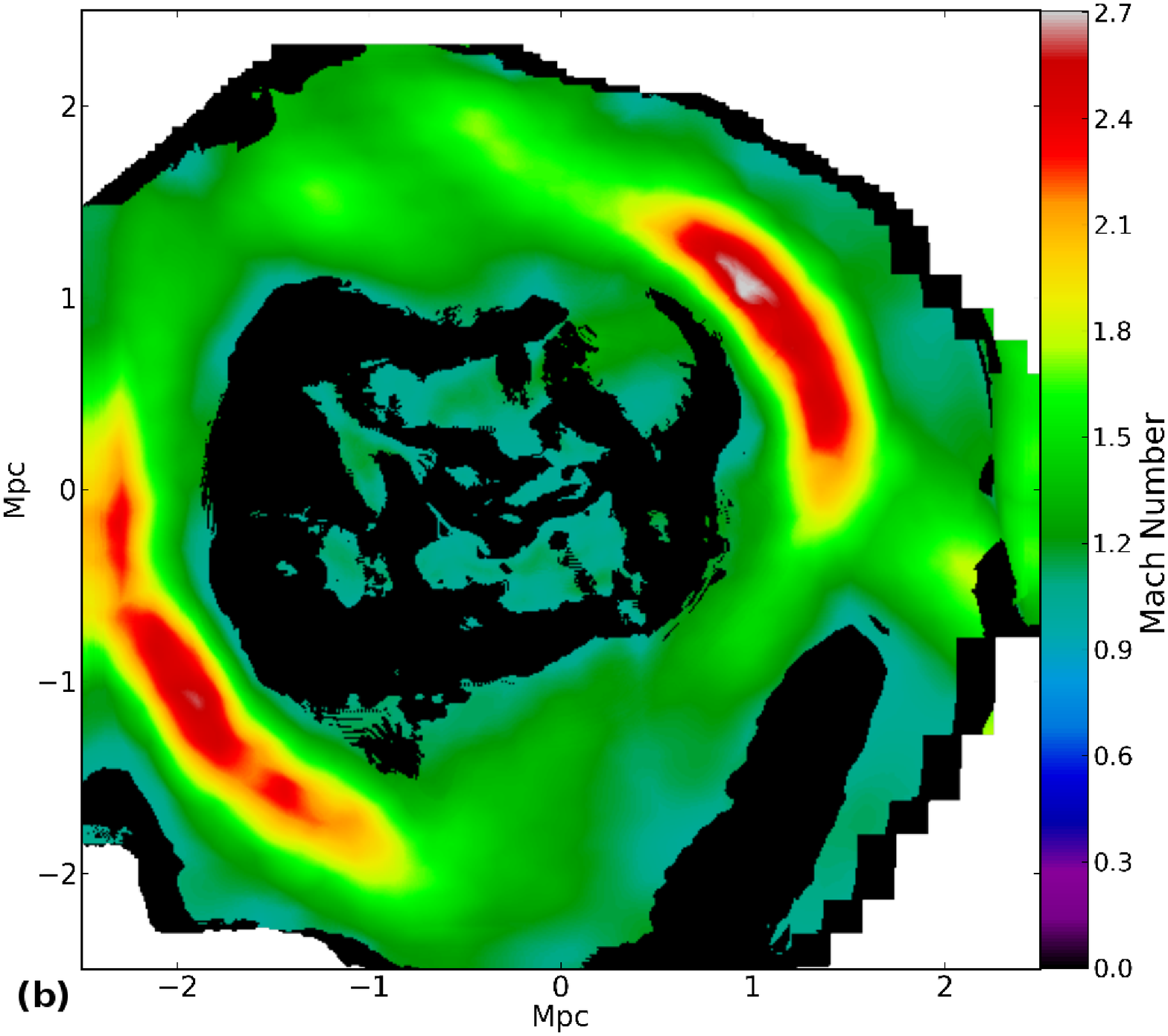,height=3.0truein}
\caption{{\bf (a)} Projected and X-ray emission weighted Mach number map obtained by running the 3-dimensional shock-finder on simulated MHD cluster (Figure~\ref{fig:simXT}). {\bf (b)} Mach number map obtained by running the 2-dimensional shock-finder on the projected, binned X-ray surface brightness and temperature maps of the simulated MHD cluster.}
\label{fig:simPM_wvt}
\end{figure*}
%------------------------------------------------------------------
Galaxy clusters are formed due to collapse of density fluctuations involving comoving scales of $\sim 10$~Mpc which contribute to their large mass ($10^{14}-10^{15}~M\odot$). Evolution of structure of the universe is mainly driven by gravity at this regime. The growth and evolution of galaxy clusters can be very accurately modeled in cosmological simulations [see \citet{borgani11}, for a review]. The adoption of adaptive mesh refinement (AMR) techniques for cosmological simulations \citep{bryan97a,bryan97b,norman99,oshea04,bryan13} has increased the spatial and temporal resolution of the cosmological hydrodynamic simulations in the region of interest. Due to this, the shock-capturing gas dynamics schemes have improved and are able to trace the galaxy cluster evolution more accurately. 

From the observations of galaxy clusters like A3667 at multi-wavelengths (X-ray, optical, radio, etc) one can only infer the cluster properties as projected on the plane of the sky. Unlike observations, current state-of-the-art AMR simulations trace the entire 3-dimensional evolution of the galaxy clusters. Hence, a combination of the observations and simulations can permit us to disentangle the contribution of different physical processes occurring in the ICM plasma. 

In this section we investigate the properties of an MHD cluster (previously reported in \citet{skillman13}) and compare some of the properties with A3667. It should be noted that this simulated cluster was never meant to represent A3667. Although there are differences between this cluster and A3667, there are also some striking similarities.

\subsection{MHD cluster} 
The simulated cluster was run using a modified version of the {\it Enzo} cosmology code \citep{bryan13}. {\it Enzo} uses block-structured AMR \citep{berger89}. The details of the simulation procedure for this cluster have been reported in \citet{skillman13} which is the same as cluster U1 in \citet{xu11}. Here, we summarize some major aspects of the simulation and then compare the simulated cluster with observations. 

In this simulation, clusters were formed from cosmological initial conditions, generated at redshift $z = 30$ from an \citet{eisenstein99} power spectrum of density fluctuations in a $\Lambda$CDM universe. The AMR criteria in this simulation are the same as in \citet{xu11}. In order to identify the shocks which ultimately accelerate the electrons that emit synchrotron radiation, the 3-dimensional shock-finder was used [as mentioned in Section~\ref{sec:sf} and \citet{skillman08}]. A minimum pre-shock temperature of $T = 10^4$~K was set since the low-density gas in the cosmological simulations is assumed to be ionized (a reasonable assumption at z < 6). Therefore, any time the pre-shock temperature is lower than $10^4$~K, the Mach number is calculated from the ratio of the post-shock temperature to $10^4$~K. The 3-dimensional shock-finder, already discussed in detail and validated \citep{skillman08,skillman11,skillman13}, has been implemented to run ``on-the-fly'' in \Enzo~ and has the unique ability to accurately identify off-axis shocks within AMR simulations and quantify their Mach number even if the shock is identified as being spread out across several cells. 

In order to estimate the synchrotron emission from the shock waves, the method of \citet{hoeft07} as adapted in \citet{skillman11} is followed. The electrons are assumed to be accelerated to a power-law distribution that is related to the Mach number from Diffusive Shock Acceleration or DSA \citep{drury83,blandford87} theory in the test-particle limit. These accelerated electrons form an extension to the thermal, Maxwellian distribution that has a power-law form and exponential cutoff related to balancing the acceleration and cooling times of the electrons. The accelerated electrons then emit in the radio through synchrotron radiation. In order to estimate the radio synchrotron emission, the magnetic fields generated in the simulations are directly used \citep{skillman13}. The magnetic field initialization used is the same method in \citet{xu09} as the original magnetic tower model proposed by \citet{li06}, and assumes the magnetic fields are from the outburst of AGNs. In this simulation, magnetic fields are injected at $z = 3$ into two proto-clusters which in turn belong to two subclusters. The merged, simulated cluster at $z = 0$ has: $R_{virial}=2.5$~Mpc, $M_{virial} = 1.9\times10^{15}M_\odot$ and $T_{virial}=10.3$~keV. This simulation models the evolution of dark matter, baryonic matter, and magnetic fields self-consistently. The simulation uses an adiabatic equation of state for gas, with the ratio of specific heat being $\gamma=5/3$, and does not include photoionization/heating or radiative cooling physics or star/galaxy feedback.

Figure~\ref{fig:simXT}(a) shows the projected X-ray emission which is calculated using bremsstrahlung and free-free emission from the Cloudy code to integrate the emission from 0.5-12.0 keV with a metallicity of $0.3 \mathcal{Z}_\odot$. The radio emission from this cluster (Figure~\ref{fig:simXT_wvt}(a)) shows a double-relic system. The primary cluster is moving from the North-West to the South-East while the secondary cluster mainly to the East. This is not the exact scenario in case of Abell 3667 or in the simulations done by \citet{roettiger99} which reproduced the observed features in simulated radio maps as in A3667. It should be noted that this simulation is only meant to be illustrative of the processes in a major merger and not reproduce the exact scenario in A3667. After the core passage, the merger shock develops and moves out in both directions towards North-West and South-East. The double-relic feature is aligned well with the central X-ray emission. The X-ray emission (Figure~\ref{fig:simXT}(a)) is also elongated along the merger axis showing its un-relaxed nature. The projected temperature (Figure~\ref{fig:simXT}(b)) is not centrally peaked as the X-ray surface brightness (a strong function of the density). The projected size of the simulated MHD cluster between the two relic positions is about $4$~Mpc, which is very close to the separation of the observed double relics in A3667.

\subsection{Comparison with the X-ray Observations}
\label{sec:compX}

Here we compare the simulated MHD cluster with the broad features in the A3667 X-ray data. In order to compare the simulated cluster properties with that of the X-ray observations of A3667, we use the WVT binning technique on the simulated cluster (same as in Section~\ref{sec:wvt}). The simulated cluster has much higher dynamic-range in X-ray surface brightness than the observations of A3667. Hence, we choose an appropriate cut in the X-ray surface brightness and converted that to X-ray counts. The X-ray counts were scaled according to the count profile of A3667 such that it represents the similar detector properties as in real {\it Chandra} observations. Since there are no background files (as in real observations), we apply the WVT-binning to the signal map only with a SNR of 50. It should be noted that the same threshold was applied to the A3667 data in Sections~\ref{sec:wvt} and \ref{sec:acb}. In this case, the noise map is just the Poisson noise from the simulated counts map. The WVT region map is used to bin the surface brightness and temperature map and then adaptively smooth them using {\it ASMOOTH} (as in Section~\ref{sec:wvt}). The WVT-binned and smoothed X-ray temperature map is shown in Figure~\ref{fig:simXT_wvt}(b). The double radio relics in this simulated cluster are located about $\sim 2$~Mpc from the cluster center. This is a close match to the distance of the observed double radio relic in A3667 from the center of the cluster. 

In order to be consistent with our shock analysis on the {\it Chandra} data, we have applied the 2-dimensional observational shock-finder to the projected, WVT-binned and adaptively-smoothed temperature map from the MHD cluster simulation. The resultant Mach number map is shown in Figure~\ref{fig:simPM_wvt}(b). Before we proceed to compare the X-ray detected shocks in the simulated MHD cluster with those from X-ray observations of A3667, we need to validate our 2-dimensional shock-finder. In order to do this comparison, we applied the original 3-dimensional shock-finder on the 3-dimensional data of the simulated MHD cluster. The resultant Mach numbers are then weighted by X-ray emission and projected onto the 2-dimensional Mach number map (see Figure~\ref{fig:simPM_wvt}(a)). Both shock-finders (Figures~\ref{fig:simPM_wvt}(a) \& (b)) identified shocks near the location of the radio relics (Figure~\ref{fig:simXT_wvt}(a)). Near the radio-relic locations, the Mach numbers for the 3-dimensional shock-finder are between $2.0 \lesssim \M \lesssim 4.5$ (Figure~\ref{fig:simPM_wvt}(a)) and the Mach numbers for the 2-dimensional shock-finder are between $2.0 \lesssim \M \lesssim 2.8$ (Figure~\ref{fig:simPM_wvt}(b)). Due to projection effects, the 2-dimensional shock-finder does not find Mach numbers as high as those detected by the 3-dimensional shock-finder at some pixel locations near the relic as well as near the outskirts. In addition, the shock structure (near the radio-relics) for the 2-dimensional shock-finder appears wider than that for 3-dimensional shock-finder. This can be attributed to the WVT binning that was applied to the projected X-ray surface brightness and temperature maps before applying the 2-dimensional shock-finder on these two maps. It should be noted that such binning procedure (see Section~\ref{sec:temp}) is required in actual X-ray observations (like A3667) to gather sufficient SNR in each region and derive temperature maps with reasonable error values. Considering the projection and binning effects, the observational 2-dimensional shock-finder has been able to detect shocks at the same locations as the 3-dimensional shock-finder. We will further validate the 2-dimensional shock-finder in section~\ref{sec:compXRS} using the Mach number distributions.

After validating our 2-dimensional shock-finder, we compare the X-ray shock features between the simulated MHD cluster with {\it Chandra} results of A3667. In the inner regions of the cluster ($\lesssim 1$~Mpc), the Mach numbers from the 2-dimensional shock-finder (Figure~\ref{fig:simPM_wvt}(b)) are $\sim 0.9-1.8$ which is consistent with the Mach number maps of A3667 (Figure~\ref{fig:Amach}). However, the agreement is only valid for the inner $\sim 1$~Mpc of A3667 due to the absence of any X-ray data extending to the radio relic locations. In the next section, we use previous radio observations by \citet{huub97} to calculate the shock Mach numbers from the region around the radio relics and then compare the combined X-ray/radio estimates of the Mach number distribution with that from the simulated MHD cluster.

\section{Shocks from Radio Observations}
\label{sec:rad}
%------------------------ Figure:- 11 -----------------------------
\begin{figure*}[t!]
\centering
\epsfig{file=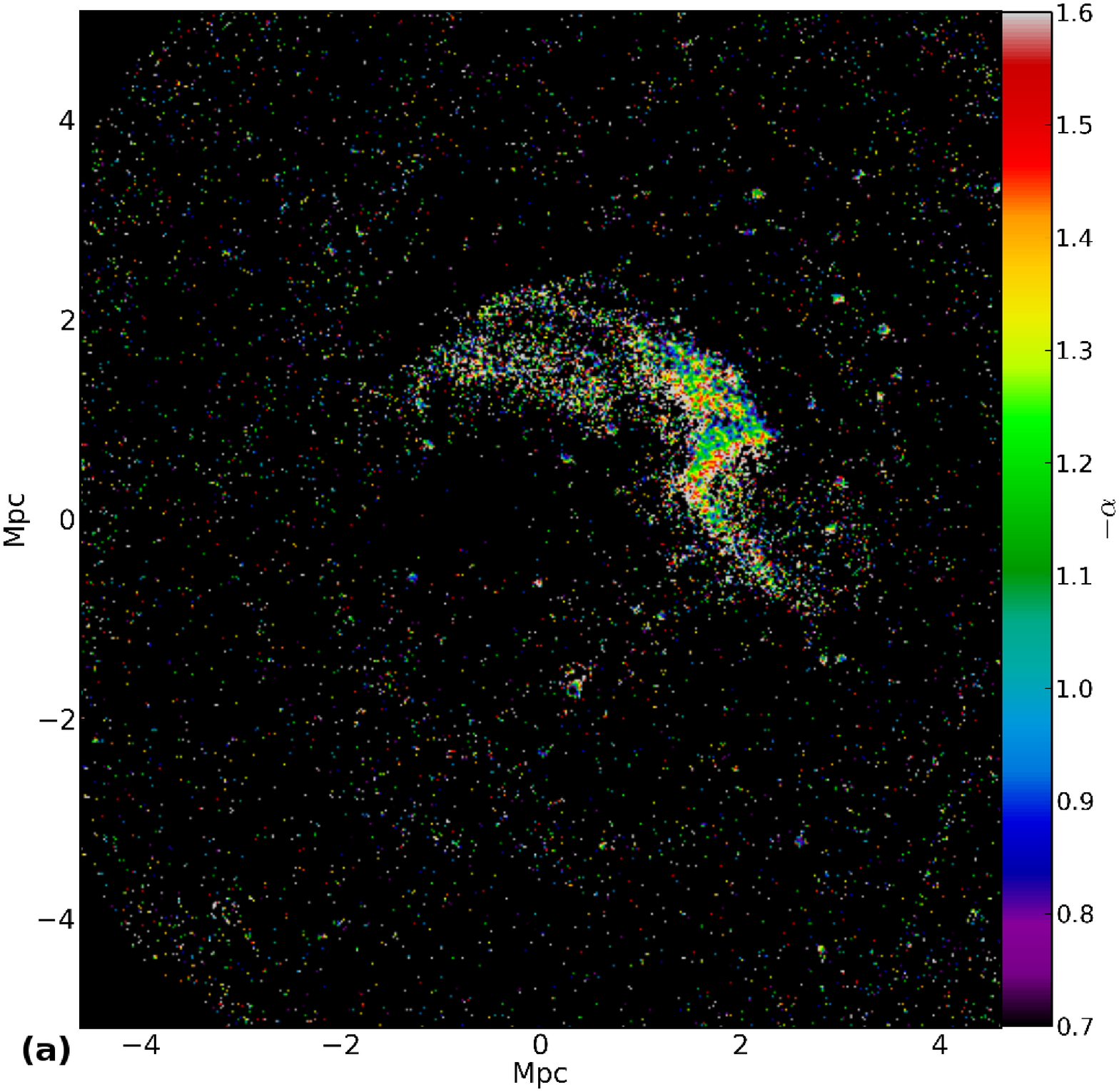,height=3.4truein}
\epsfig{file=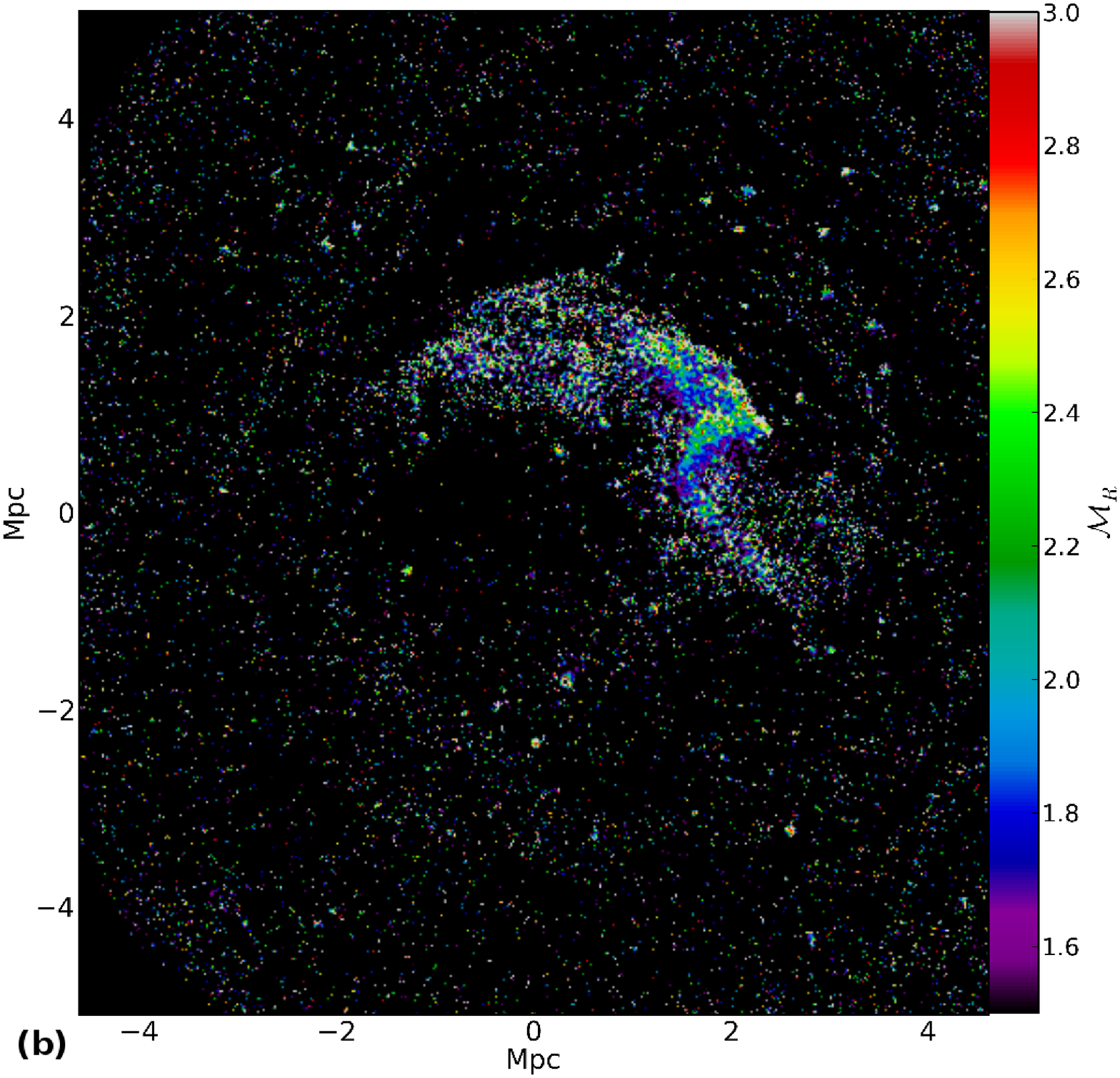,height=3.4truein}
\caption{{\bf (a)} Radio Spectral index map as obtained from the maps of radio observations at 13 and 20 cm [Huub R\"{o}ttgering (private communication)]. The observations were carried out with the ATCA \citep{huub97}. {\bf (b)} Radio Mach number map obtained from the spectral index map following \citet{rosswog07}.}
\label{fig:Arad}
\end{figure*}
%------------------------------------------------------------------

Over the past decade, sensitive radio observations have increased our knowledge of the properties of the non-thermal components of the ICM. Several clusters have been detected in the radio wavelengths showing diffuse synchrotron emission in the central (radio halo) and peripheral (radio relic) cluster regions \citep{feretti12}. Rotation Measure studies in the radio bands reveal the existence of large scale magnetic fields in the ICM which is amplified by the 
shocks \citep{luigi12}. The radio relics are associated with the shocks and have moderately polarized radio emission with spectral indices of $\alpha \sim 1-2$ for surface brightness $ S \propto \nu^{-\alpha} $ \citep{huub97,bagchi06,weeren11b} . This spectral shape suggests that the electrons are recently shock-accelerated \citep{blandford87}. These electrons undergo diffusive shock acceleration or DSA \citep{drury83,schure12}, which is a first-order Fermi mechanism where electrons are accelerated by reflecting off magnetic field perturbations created by plasma effects in shock waves.  

The MHD cluster discussed in section~\ref{sec:sim} used magnetohydrodynamic evolution of a galaxy cluster from cosmological initial conditions to model radio relics \citep{skillman13}. Shocks are located using the 3-dimensional shock-finder \citep{skillman08} as discussed in Section~\ref{sec:sf} and models of cosmic ray electron acceleration are applied. The thermodynamic properties of the radio-emitting plasma are utilized to make synthetic radio maps for this simulated cluster which shows a double relic system \citep{skillman13}.

So far in this paper, we compared the shocks in A3667 with the simulated MHD cluster (as discussed in Section~\ref{sec:sim}) within the inner $\sim 1$~Mpc region of A3667. In order to study the shocks beyond the $\sim 1$~Mpc region, we now include the radio observations from \citet{huub97} as shown in Figure~\ref{fig:rgb}(a). The radio relics extend up to $\sim 1.8$~Mpc from the center of the cluster which is similar to the distance of the relic in the simulated MHD cluster from its center. 

The radio observations were made at 13 and 21~cm wavelengths with the ATCA. The details of the observations are discussed in \citet{huub97}. We have used the radio maps at the two frequencies shared by Huub R\"{o}ttgering in a private communication.  The restoring beam size or resolution of these images are $18 \times 18$~arcsec. From these images, a spectral index map was created between these two frequencies with a Python\footnote{http://www.python.org/} script. Figure \ref{fig:Arad}(a) shows the spectral index map. The spectral index map is similar to the one reported in \citet{huub97}. In this map, the North-West relic is clearly visible but not the South-East counterpart. This is due to the fact that the 13cm observations from the ATCA has the South-East relic at the very edge of the image which is also dominated by high RMS noise. Since the region around the South-East relic has a very low Signal-to-Noise ratio in the higher frequency map, it is not significant in the spectral index map. 

In A3667, as we are viewing the relic ``edge-on'', the radio spectral index ($\alpha$) should be sensitive to the prompt emission from the shock front \citep{skillman13} and is given by $\alpha=\alpha_{prompt} = (1-s)/2$, where $s$ is the spectral index of the accelerated electrons given by $n_e(E) \propto E^{-s}$ \citep{hoeft07}. The theory of diffusive shock acceleration (DSA) for planar shocks, at the linear test-particle regime, predicts that this radio spectral index is related to the shock Mach number by  \citep{blandford87,hoeft07, ogrean13}:
\begin{equation}\label{eq:radmach}
\M^2=\frac{2\alpha-3}{2\alpha+1}
\end{equation}
where $\alpha$ is the radio spectral index ($S_R \propto \nu^\alpha$). Following the above equation we have constructed the Mach number map for the radio data shown in Figure \ref{fig:Arad} (b). Mach numbers are found to be $\sim 2.6$ near the outer edge of the North-West relic. Previous studies of A3667 with XMM-Newton \citep{finoguenov10} derived the Mach number near the relic to be $2.4\pm0.77$. This is consistent with our results. Alternatively, if the relic is viewed ``face-on'' the radio spectral index is sensitive to cumulative spectral index ($\alpha=\alpha_{integrated}=-s/2$) due to emission from electrons over the entire lifetime. Hence, the corresponding Mach number is given by $\M^2=(\alpha+1)/(\alpha-1)$ \citep{pinzke13,skillman13}. Although the relic in A3667 is viewed ``edge-on'', the exact estimate of the outer edge of the radio-relic (or the leading edge of the shock front)  can only be estimated with respect to the dynamic range in the radio images. Hence, there is a possibility that our Mach number map (Figure \ref{fig:Arad} (b)) includes some aged electron population. Here, we also estimate the Mach number derived from the integrated spectral index at the radio-relic location of A3667. The integrated spectral index around relic region for A3667 is $\alpha_{integrated}= -1.47$ which gives us a Mach number of $2.29$. This is also in agreement with the XMM results of \citet{finoguenov10}.

\section{Comparison of the Simulations with X-ray and Radio Observations}
\label{sec:compXRS}
%------------------------ Figure:- 12 -----------------------------
\begin{figure*}[t!]
\centering
\epsfig{file=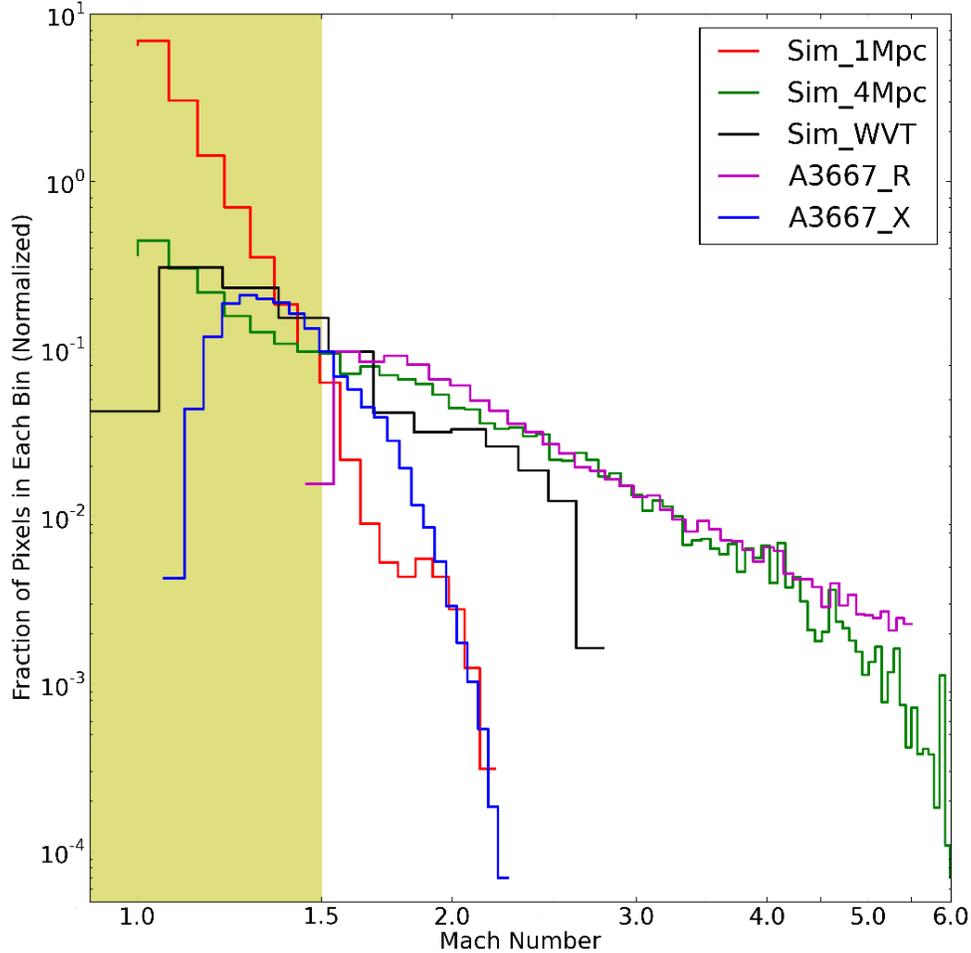,height=5truein}
\caption{Mach number distributions from X-ray and radio data along with the simulated MHD cluster: Mach number distributions from the 3-dimensional shock-finder are shown for the inner $\sim 1$~Mpc region of the MHD cluster (in red) and the entire $\sim 4$~Mpc region including the location of the double radio-relics (in green). The black line corresponds to the Mach number distributions from the observational shock-finder applied to the projected and WVT-binned temperature map of the simulated MHD cluster. The blue line shows the Mach number distribution from the observational shock-finder applied to the ACB temperature map of A3667. The magenta line corresponds to the radio Mach number distribution derived from the ATCA observations of A3667 \citep{huub97}. The region below Mach Number 1.5 is shaded in yellow as all the curves are normalized for this Mach number value. }
\label{fig:machhist}
\end{figure*}
%------------------------------------------------------------------
In section~\ref{sec:compX}, we have already shown that our observational 2-dimensional shock-finder identifies shocks at the same locations as the original 3-dimensional shock-finder \citep{skillman08}. However, the Mach number values are higher at some pixel locations near the shocks for 3-dimensional shock-finder than in the 2-dimensional counterpart. Here, we extend the validation of the 2-dimensional shock-finder by comparing the Mach number distribution calculated from the WVT-binned projected temperature maps of the simulated MHD cluster (Figure~\ref{fig:simPM_wvt}(b)) with the original 3-dimensional shock-finder that is used in the cosmological simulations. Then we will proceed to compare the observational shock-finder results on the A3667 X-ray and the Mach number distribution from the radio data with the 3-dimensional shock-finder results from the simulated MHD cluster. In order to guide the reader through the interpretation in this section, we refer to Figure~\ref{fig:machhist} which summarizes all the relevant results to be discussed in this section. 

Since the A3667 X-ray data do not extend near the location of the double radio relics, we have used two Mach number distributions from the 3-dimensional shock-finder: a) {\it Sim-1Mpc} curve -- for the inner $\sim 1$~Mpc region of the MHD cluster (red line in Figure~\ref{fig:machhist}) and b) {\it Sim-4Mpc} curve -- including the entire MHD cluster of $\sim 4$~Mpc size (green line in Figure~\ref{fig:machhist}) including the double radio relics. The Mach number values in the {\it Sim-1Mpc} curve only extends to $\mathcal{M} \lesssim 2$. This is consistent with the fact that the shocks at the inner part of the cluster are expected to have lower Mach numbers. On the other hand, the {\it Sim-4Mpc} curve includes the outskirts of the cluster and traces shocks with higher Mach number $\lesssim 6$, same as in Figure~\ref{fig:simPM_wvt}(a). It should be noted that results from the 3-dimensional shock-finder do not suffer from the projection effects and can find shocks with higher Mach number than the observational shock-finder which operates on the 2-dimensional projected surface brightness and temperature maps. 

Next, we compare the {\it Sim-1Mpc} and {\it Sim-4Mpc} curves with the {\it Sim-WVT} curve (black line in Figure~\ref{fig:machhist}) -- the result from the observational shock-finder operating on the projected and WVT-binned temperature map of the simulated MHD cluster. We should note that different Mach number distributions shown in Figure~\ref{fig:machhist} are normalized with the {\it Sim-WVT} curve at Mach number value of 1.5. Due to this normalization the agreement between the slopes of different Mach number distribution is more pronounced in the region above Mach of 1.5. Hence, the region below Mach of 1.5 is shaded in the Figure~\ref{fig:machhist} and we have less confidence in distributions for these Mach numbers. The slope of the Mach number distribution (in black) matches with the {\it Sim-4Mpc} curve. However, the Mach number values in {\it Sim-WVT} curve are restricted to $\mathcal{M} \lesssim 3$. This can be due to two reasons. Firstly, in Section~\ref{sec:sim} we have used an appropriate cut on the X-ray surface brightness of the simulated cluster in order to compare with A3667 X-ray results. This cut has restricted the field-of-view of the simulated MHD cluster to only the regions $\lesssim 4$~Mpc from the center of the cluster. Hence, the {\it Sim-4Mpc} curve may include regions at larger radii from the center of the cluster than in the {\it Sim-WVT} curve. Secondly, the {\it Sim-WVT} curve also suffers from projection effects, which can make strong shocks to appear weaker. Considering the above two limitations, we can infer that the results from the 3-dimensional shock-finder are in agreement with that of the observational shock-finder and successfully validates the latter. 

We now proceed to compare the {\it Sim-1Mpc} curve with the results of the observational shock-finder on A3667 X-ray data. The {\it A3667-X} curve (blue line in Figure~\ref{fig:machhist}) corresponds to the Mach number distribution from the observational shock-finder when applied on the X-ray surface brightness and ACB temperature map of A3667 (Figure~\ref{fig:Atemp_acb}(a)). It is relevant to note here that the combined {\it Chandra} observations of A3667 are only limited to the central $\sim 1$~Mpc from the center of the cluster and do not extend to the location of the double radio relics. This is similar to the case of {\it Sim-1Mpc} curve. Figure~\ref{fig:machhist} shows that the slopes of these two Mach number distributions ({\it A3667-X} and {\it Sim-1Mpc} curves) are well in agreement. The Mach number values in {\it A3667-X} curve are restricted to $\lesssim 2$ as in {\it Sim-1Mpc} curve. Also, the slopes of both these distributions match very well considering the fact that the {\it A3667-X} curve is influenced by the binning with ACB circles (shown in Figure~\ref{fig:Atemp_acb}(b)). 

Now, we include the A3667 radio data which extends to the location of the double relics (Figure~\ref{fig:rgb}(a)) and traces shocks at the outskirts of the A3667 cluster ($\sim 1.8$~Mpc from the center of the cluster). This is similar to the case of {\it Sim-4Mpc} curve. The {\it A3667-R} curve (magenta line in Figure~\ref{fig:machhist}) corresponds to the Mach number distribution from the radio Mach number map (Figure~\ref{fig:Arad}(b)). The radio Mach number values in {\it A3667-R} curve extends to higher Mach numbers ($\lesssim 6$) as in {\it Sim-4Mpc} curve. Moreover, the slope of the two distributions ({\it A3667-R} and {\it Sim-4Mpc} curves) are in excellent agreement. It should be again noted that the simulated MHD cluster was never meant to reproduce the morphology of the A3667 observations and is only meant to illustrate the processes in a major merger. The similarity in the Mach number distributions between A3667 and the MHD cluster, irrespective of the morphology or the dynamics, might hint at the fact that all merging clusters have similar shock statistics. 

From the results discussed in this section, we have three major conclusions. Firstly, we can conclude that the shocks in a real observed galaxy cluster like A3667 can be reproduced in the current state-of-the-art MHD AMR simulations [e.g. \citet{skillman13}] due to accurate modeling of the physical processes in the ICM. Secondly, we have demonstrated the need to combine the observations at X-ray and radio wavelengths to analyze the shocks in a galaxy cluster (like A3667) accurately. This is a key result from our work. Finally, we have validated a 2-dimensional observational shock-finding algorithm that can now be used to detect shocks in other X-ray data on galaxy clusters.

\section{Structure of the Cold Front}
\label{sec:cf}
%------------------------ Figure:- 13 -----------------------------
\begin{figure*}[t!]
\centering
\epsfig{file=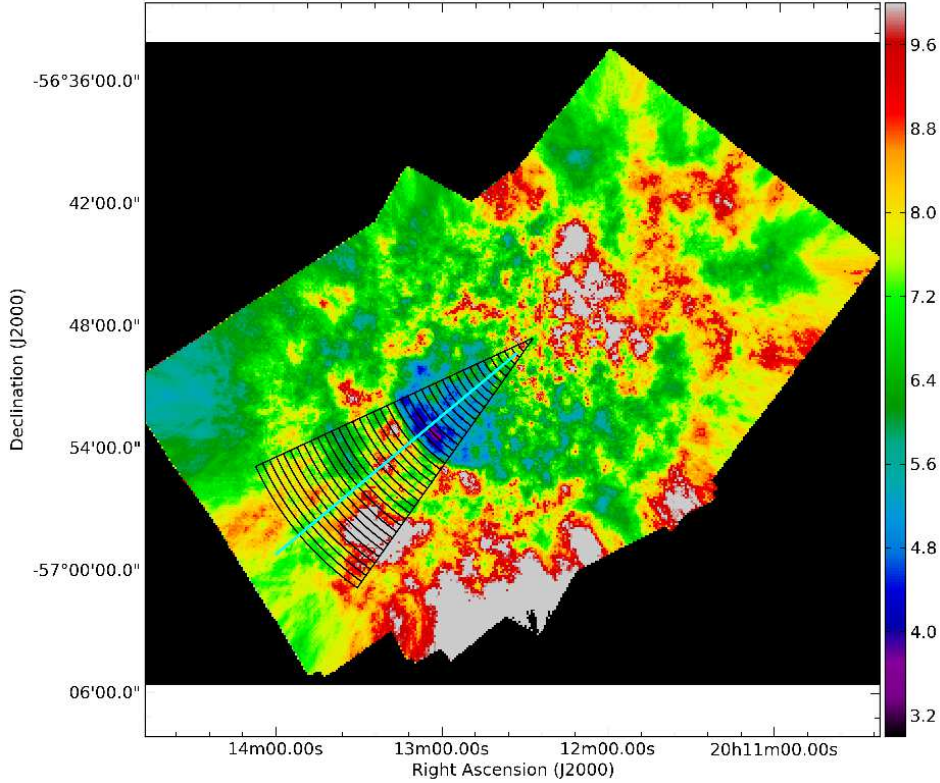,height=4.2truein}
\caption{The annular regions, over which the cold front temperatures have been extracted, are overlayed on top of the ACB temperature map (in black). Also shown the line (in cyan) along which a slice is extracted from the ACB temperature map for higher resolution data across the cold front. The ACB temperature map is same as in Figure~\ref{fig:Atemp_acb}. The colorbar shows the temperature values in keV.}
\label{fig:A_overlay}
\end{figure*}
%------------------------------------------------------------------

\renewcommand{\tabcolsep}{3pt}
\begin{deluxetable*}{ccccccccccc}
\tablewidth{0pt}
\tabletypesize{\small}
\tablecaption{Cold Front Analysis}
\tablecolumns{11}
\tablehead{\colhead{Region} &  \colhead{$\Delta R$} & \colhead{$kT_x$ } & \colhead{$\left(\frac{kT_{in}}{kT_{out}}\right)$} & \colhead{$n_e$} & \colhead{$\left(\frac{n_e^{in}}{n_e^{out}}\right)$} & \colhead{$\left(\frac{P^{in}}{P^{out}}\right)$} & \colhead{$\left(\frac{S_E^{in}}{S_E^{out}}\right)$} & \colhead{$\lambda_e$} & \colhead{$\lambda_e^{in \rightarrow  out}$} & \colhead{$\lambda_e^{out \rightarrow in}$}} 
\startdata
~            & kpc       &  keV           & ~          & $10^{-3}~cm^{-3}$ & ~         & ~         & ~         & kpc     & kpc     & kpc    \\  \hline 
In (Annuli)  & {\bf $24$}   &  $4.39\pm0.01$ & $0.58$     & $3.42\pm0.01$   & $2.46$    & $0.97$    & $0.38$    & $1.73$  & $14.73$ & $2.18$ \\
Out (Annuli) & {\bf $\pm12$}&  $7.48\pm0.13$ & $\pm 0.06$ & $1.39\pm0.01$   & $\pm0.01$ & $\pm0.09$ & $\pm0.03$ & $12.32$ &  ~      & ~   \\ \hline
In (ACB)     & {\bf $26$}   &  $4.58\pm0.31$ & $0.65$     &  --             & --        & $0.95$    & $0.65$    & $1.88$  & $11.61$ & $2.22$ \\
Out (ACB)    & {\bf $\pm2$} &  $6.99\pm0.65$ & $\pm0.09$  &  --             & --        & $\pm0.23$ & $\pm0.12$ & $10.76$ &  ~      & ~ 
\enddata
\tablecomments{Details of the thermodynamic quantities across the Cold Front. The treatment is similar to previous work by \citet{vikhlinin01b}. $\frac{P^{in}}{P^{out}}=\left(\frac{n_e^{in}}{n_e^{out}}\right)\left( \frac{kT_{in}}{kT_{out}}\right)$ and $\frac{S_E^{in}}{S_E^{out}}=\left(\frac{n_e^{out}}{n_e^{in}}\right)^{2/3} \left(\frac{kT_{in}}{kT_{out}}\right)$.}
\label{tab:cf}
\end{deluxetable*}

%------------------------ Figure:- 14 -----------------------------
\begin{figure*}[t!]
\centering
\epsfig{file=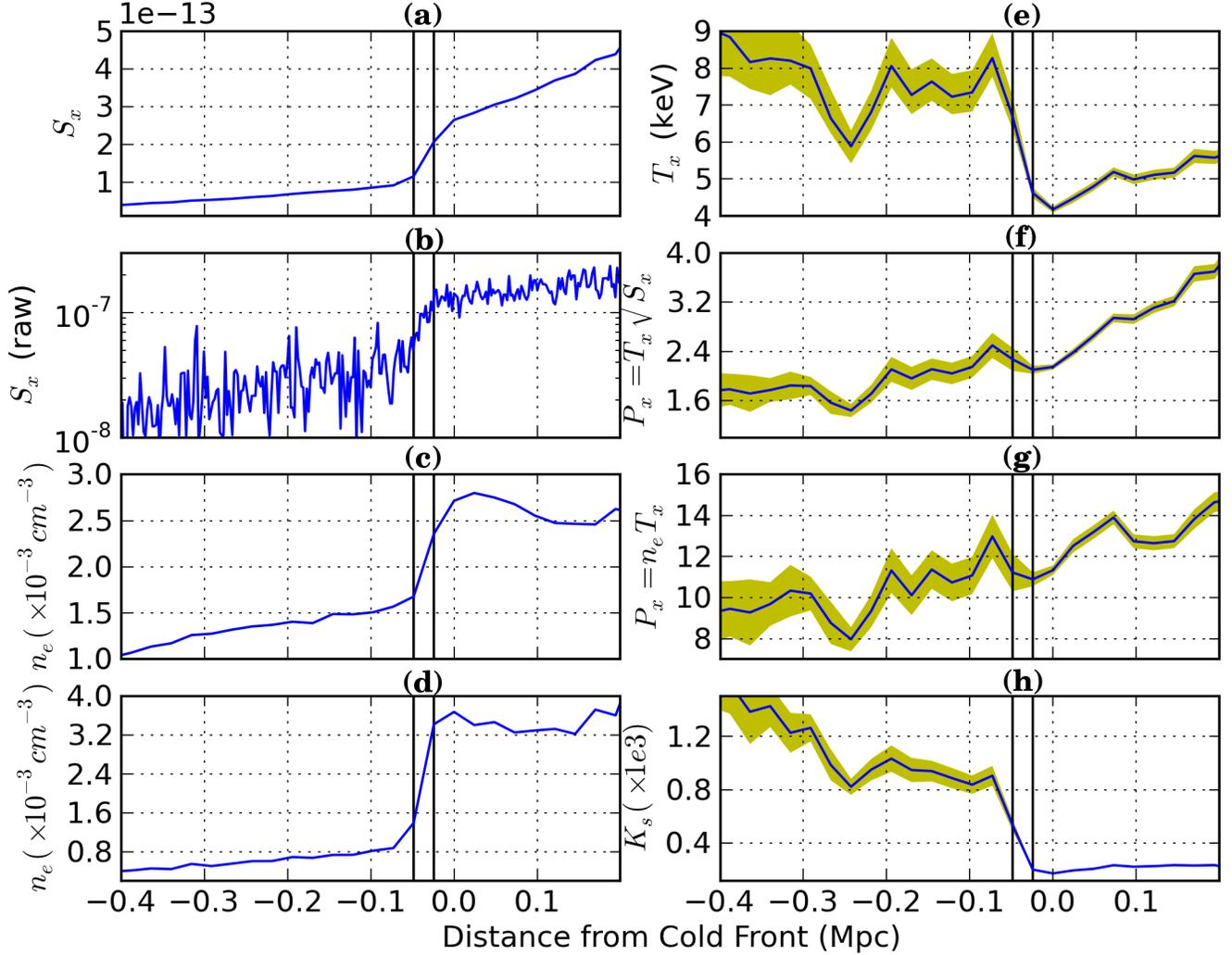,height=5.5truein}
\caption{Eight panels showing temperature, surface brightness, pressure and entropy variation across the cold front. This plot corresponds to the annular sectors (Figure~\ref{fig:A_overlay}) extracted across the cold front [as discussed in section~\ref{sec:cf}]. {\bf (a)} The X-ray surface brightness averaged over each annular sectors is shown. The vertical lines show the approximate width of the cold front. The error-bars in the X-ray surface brightness are also shown but they are nearly the width of the curve. {\bf (b)} Pixel-by-pixel X-ray surface brightness along the line (shown in cyan: Figure~\ref{fig:A_overlay}). {\bf (c)} The electron density profile calculated from the APEC normalization (equation~\ref{eq:apec_norm}). {\bf (d)} Deprojected electron density profile calculated from Equation~\ref{eq:apec_dp}. {\bf (e)} The projected temperature as fitted from the extracted spectra over each annular sectors. The $1-\sigma$~errorbars are also shown. {\bf (f)} The pseudo-pressure across the cold front ($P=\sqrt{S_x}T_x$). {\bf (g)} The pressure (in $1.6\times10^{-13}$~Pa) across the cold front ($P=n_eT_x$) as derived from the deprojected density. {\bf (h)} The entropy across the cold front ($K_S=n_e^{-2/3}T_x$) as derived from the deprojected density. See the text for discussion.}
\label{fig:cf_pie}
\end{figure*}
%------------------------------------------------------------------

%------------------------ Figure:- 15 -----------------------------
\begin{figure}[t!]
\centering
\epsfig{file=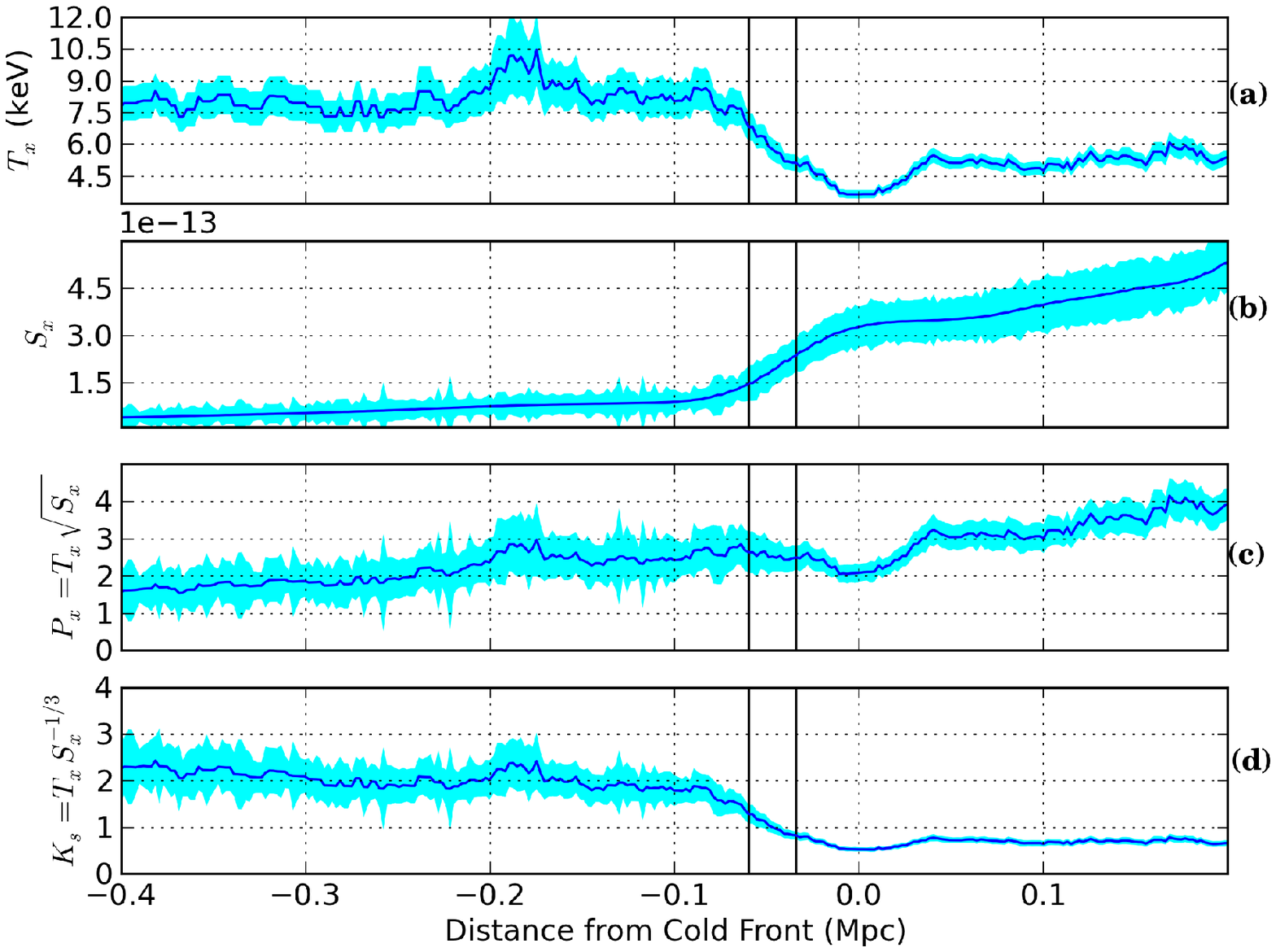,height=2.6truein}
\caption{Four panels showing temperature, surface brightness, pressure and entropy variation across the cold front along the 'cyan' line in Figure~\ref{fig:A_overlay}. {\bf (a)} Temperature along the line derived from the ACB map (Figure~\ref{fig:Atemp_acb}(a)). {\bf (b)} X-ray surface brightness profile along that line. {\bf (c)} Pseudo-pressure profile: $P=\sqrt{S_x}T_x$. {\bf (d)} Pseudo-entropy profile : $K_S=S_x^{-1/3}T_x$. See the text for discussion.}
\label{fig:cf_acb}
\end{figure}
%------------------------------------------------------------------

Cold fronts in galaxy clusters were first confirmed in Abell 2142 \citep{markevitch00} and in Abell 3667 \citep{vikhlinin01b} from the high resolution {\it Chandra} observations. Cold fronts are seen as an edge-like feature in the X-ray surface brightness map, but the density and temperature jumps across the cold front do not follow the Rankine-Hugoniot shock jump conditions. Moreover, the pressure is continuous across the cold front which confirms that the cold fronts are contact-discontinuities and not shocks. Several other galaxy clusters are also found to have a single [e.g. 1ES0657-558, Abell 1201 \citep{owers09b}] or multiple [e.g. Abell 521 \citep{bourdin13}] cold fronts in {\it Chandra} and XMM-Newton observations. Cosmological hydrodynamic simulations show that origin of cold fronts during a cluster merger can happen in multiple ways \citep{mathis05,poole06,ascasibar06,hallman10,zuhone13}. Cold fronts can be classified broadly as: a) ``remnant core''- where the discontinuity occurs between infalling subcluster's cool-core and hot ambient ICM of the main cluster and b) ``sloshing''- where the discontinuity occurs at the interface of the higher entropy gas at the outer region of the cluster and the cool, dense gas which is displaced from the cluster core due to some disturbance [see \citet{markevitch07} for a review].

Here, we study the cold front in A3667. In a previous study, \citet{vikhlinin01b} used a single {\it Chandra} observation (exposure time $\sim 49$~ksec) to derive the thermodynamic properties across the cold front in A3667 and constrain the width of the front. Moreover, their analysis suggested that the transport processes across the front should be suppressed to maintain the sharpness of the front. In addition, their analysis suggested the existence of a possible bow-shock about $\sim 350$~kpc away from the cold front. 
  
 In order to further investigate the cold front in A3667 with the combined {\it Chandra} observations of $\sim 447$~ksec, we created sectors of concentric annuli between position angle $215\deg$ and $245\deg$ (same as in \citet{vikhlinin01b}). The center of the annuli were located near the center of the cluster. We created 40 annular sectors with varying radial width such that each sector contain sufficient SNR in order to fit temperature to each extracted spectrum from each region. Figure~\ref{fig:A_overlay} shows the exact location of these annular sectors with respect to the ACB temperature map. Moreover, the spectra from different observations corresponding to the same region were combined following the same method as described before in the section~\ref{sec:wvt}. Since these annuli have to be of certain radial width in order to have sufficient SNR, the exact width of the cold front can be overestimated by this method. On top of this there can be projection effects which can further increase the width of the cold front. Hence in order to get a higher resolution estimate of the cold front width, we also used the ACB temperature map and derived a temperature profile along the line shown in cyan in Figure~\ref{fig:A_overlay}.

\subsection{Thermodynamic Properties across the Cold Front}

Here we discuss the profiles of thermodynamic quantities like density, temperature, pressure, and entropy across the cold front in A3667. Figure~\ref{fig:cf_pie} shows the profiles of the thermodynamic properties for the annular sectors as shown in Figure~\ref{fig:A_overlay}. The profile of average X-ray surface brightness over each annular sector (Figure~\ref{fig:cf_pie}a) shows a jump by a factor of $\sim 2$ over the width of {\bf $24$~kpc} (distance between the centers of the adjacent annuli across the front edge, as shown in Figure~\ref{fig:A_overlay}). Since the adjacent annular bins are of same radial widths, the error in the width of the cold front can be calculated as half of the width of the annular bin. Hence, the cold front width can be estimated as {\bf $\Delta R = 24\pm12$~kpc}. However, the unsmoothed surface brightness profile (Figure~\ref{fig:cf_pie}b) shows a jump by a factor of $\sim 2$ over $12$~kpc width, which is consistent with the width of the cold front ($\Delta R$) as estimated above. 

Inside {\it XSPEC}, the APEC normalization ($\eta_{APEC}$) from spectral fitting can be used to derive the electron density over each of the concentric annular sectors. Following \citet{walker13} and \citet{henry04}, we calculated the electron density ($n_e$) as:
\begin{equation}\label{eq:apec_norm}
n_e = 2.18\times10^{-5}h_{70}^{-1/2}\sqrt{\frac{\eta_{APEC}}{V(Mpc^3)}} D_A(Mpc)(1+z)
\end{equation}
The volume of each annular sector is given by $V \approx (4/3)D_A^3\Omega(\theta_{out}^2 -\theta_{in}^2)^{1/2}$ where $\Omega$ is the solid area of the annular sector while $\theta_{out},\theta_{in}$ are the outer and inner angular distances to the edges of the annular sectors from the center of the cluster \citep{walker13}. 
Figure~\ref{fig:cf_pie}c shows the electron density from the above calculation (equation~\ref{eq:apec_norm}). The electron density shows a jump of $\gtrsim 1.7$ over the $24$~kpc width of the front (shown as two vertical lines in all the plots of Figure~\ref{fig:cf_pie}. The electron density of the annular sector is still a projected quantity on the plane of the sky. In order to estimate the real electron densities we apply spherical deprojection technique \citep{kriss83,wong08,blanton09}. Equation A1 in \citet{kriss83} presents a relation to estimate the emission density $C_{ij}$(counts~$s^{-1} cm^{-3})$. In order to convert the emission density to the electron density, we follow \citet{cavagnolo09} and \citet{donahue06}:
\begin{equation}\label{eq:apec_dp}
n_e(r)=\sqrt{\frac{(n_e/n_p)4\pi[D_A(1+z)]^2C_{ij}(r)\eta_{APEC}(r)}{10^{-14}f(r)}}
\end{equation}
where $r$ is the distance from the center of the cluster and $f(r)$ is interpolated spectroscopic count rate. 
Figure~\ref{fig:cf_pie}d shows the deprojected electron density from the above calculation (equation~\ref{eq:apec_dp}). The sharp jump in the electron density by a factor of $2.46\pm0.01$ (see Table~\ref{tab:cf}) over the cold front width ($\Delta R$) is evident from the plot. It should be noted that the deprojected electron density shows a much higher jump across the cold front than the projected $n_e$ (Figure~\ref{fig:cf_pie}c). 

Figure~\ref{fig:cf_pie}e shows the fitted temperature for each of the 40 annular sectors. In order to obtain a deprojected temperature profile, we would have needed much larger annuli in order to achieve sufficient counts to fit multi-component APEC for the deprojection analysis. The deprojection of temperature will have a much larger uncertainty over the actual width of the cold front. Hence, we have not carried out a temperature deprojection. The projected temperature jump across $\Delta R$ is about $0.58\pm0.06$ (from inside to outside). Figure~\ref{fig:cf_pie}f shows the pseudo-pressure profile ($P=T_x\sqrt{S_x}$) across the cold front. This profile shows that there is no jump in the pressure across the cold front. Figure~\ref{fig:cf_pie}g shows the pressure (product of deprojected $n_e$ and projected $T_x$: $P=n_eT_x$) across the cold front. A cold front is a contact discontinuity and there should not be any jump in the pressure across the cold front. The pressure jump across the cold front is calculated to be $0.97\pm0.09$ (Table~\ref{tab:cf}) between the regions inside the front to the outside. This continuity in pressure confirms the cold front in A3667 to be a contact discontinuity. Figure~\ref{fig:cf_pie}h shows the entropy profile ($K_S=n_e^{-2/3}T_x$) with the deprojected electron density. The jump in the entropy is calculated to be $0.38\pm0.03$ across the cold front (inside to outside). All these thermodynamic parameters are evaluated and tabulated in Table~\ref{tab:cf}. 

Previous study by \citet{vikhlinin01b} derived the temperature jump across the cold front to be $0.53\pm0.06$ and density jump to be $3.9\pm0.83$. In a later study by \citet{owers09b}, they found the temperature jump across the cold front to be $0.45\pm0.08$ and the corresponding density jump to be $2.6\pm0.2$. The temperature jump from our analysis ($T_{in}/T_{out}=0.58\pm0.06$) is consistent with the previous two studies. The density jump across the front is consistent with that in \citet{owers09b}. However, the density jump reported in \citet{vikhlinin01b} is much higher than our analysis. \citet{vikhlinin01b} derived the density outside the cold front from their $\beta$-model fit to the X-ray surface brightness at regions extending beyond $\sim 70$~kpc from the cold front. In our analysis, the density outside the cold front is derived within $\sim 12.14$~kpc from the edge of the cold front. Hence, we obtained a higher value of density outside the cold front than in \citet{vikhlinin01b} which resulted in lower density jump. 

\citet{vikhlinin01b} derived the width of the cold front to be $2$~kpc based on the X-ray surface brightness jump only. The temperature jump was derived over a much larger width $\gtrsim 40$~kpc (see Figure 4b in \citet{vikhlinin01b}). In our analysis, we find the width of the cold front to be {\bf $\Delta R = 24\pm12$~kpc} from both the density and temperature profiles. \citet{owers09b} derived the temperature and density jumps across the cold front within a region of $\sim 70$~kpc from the cold front. Here, the combined {\it Chandra} observations allowed us to derive a higher resolution WVT and ACB temperature maps with lower error estimates. Hence, we are able to derive a much tighter constrain on the width of the cold front (based on both temperature and density jumps) than in any previous studies. Due to the sizes ($24$~kpc) of the annular sectors (Figure~\ref{fig:A_overlay}), it is possible that the cold front is still not resolved and our estimate of the cold front width is an upper-limit.

The concentric annuli have a significant width so that there are sufficient counts for a temperature fit. In order to check whether the width of the cold front inferred from this above analysis is over-estimated, we used the ACB temperature map (Figure~\ref{fig:A_overlay}). We have used a line slice (cyan line in Figure~\ref{fig:A_overlay}) through the ACB temperature map and smoothed X-ray surface brightness map to derive the temperature, surface brightness, pressure and entropy across the cold front. Figure~\ref{fig:cf_acb} shows the profiles across the cold front using the ACB temperature map. The values of the thermodynamic parameters are tabulated in Table~\ref{tab:cf}. The width of the cold front is estimated as {\bf $\Delta R = 26\pm2$}. The temperature jump across the cold front is $0.65\pm0.09$ over this width. The entropy profile in Figure~\ref{fig:cf_acb} gives a jump of $0.65\pm0.12$. The pressure from the ACB temperature map is continuous across the cold front with a jump of $0.95\pm0.23$ [see Table~\ref{tab:cf}]. Although the ACB temperature map has higher resolution, the ACB circles are not designed to trace the cold front. However, the continuity in pressure from these projected profiles across the cold front again confirms the same to be a contact discontinuity.

\subsection{Stability of the Cold Front - Constraints on Thermal Conduction}

In this section, we investigate the effect of the transport processes in maintaining the sharpness of the cold front in A3667. The sharpness of the cold front in A3667 suggests that the heat transport processes across the front are suppressed. Previously, there were several studies to investigate the stability of the cold fronts \citep{vikhlinin01b,asai04,churazov04,asai05,xiang07,zuhone13}. 

{\it (i) Diffusion Across the Cold Front: } In order to understand the transport processes across the cold front, we have compared the width of the cold front (as calculated in the previous section) with the Coulomb mean free path of the electrons and protons in the plasma on either side of the cold front. The Coulomb mean free path of the electrons ($\lambda_e$) at a location with electron density $n_e$ and temperature $T_x$ is given by \citep{vikhlinin01b,spitzer62}:
\begin{equation}\label{eq:col_mfp}
\lambda_e = 15 kpc \left(\frac{T_x}{7~keV}\right)^2 \left(\frac{n_e}{10^{-3}~cm^{-3}}\right)^{-1}
\end{equation}

From the equation~\ref{eq:col_mfp}, we can deduce the mean free paths of thermal particles in the gas on inside ($\lambda_e^{in}$) and outside ($\lambda_e^{out}$) of the cold front. In order to estimate the Coulomb diffusion of the particles traveling across the cold front, we also need the mean free paths of the particles from one side of the front crossing into the gas on the other side. They are given by $\lambda_e^{in \rightarrow out}$ and $\lambda_e^{out \rightarrow in}$ \citep{spitzer62}:
\begin{equation}\label{eq:cross_mfp1}
\lambda_e^{in \rightarrow out}  =  \lambda_e^{out} \frac{T_x^{in}}{T_x^{out}} \frac{\mathcal{G}(1)}{\mathcal{G}(\sqrt{T_x^{in}/T_x^{out}})}
\end{equation}
\begin{equation}\label{eq:cross_mfp2}
\lambda_e^{out \rightarrow in} = \lambda_e^{in} \frac{T_x^{out}}{T_x^{in}} \frac{\mathcal{G}(1)}{\mathcal{G}(\sqrt{T_x^{out}/T_x^{in}})} 
\end{equation}
where $\mathcal{G}(x)=[\Phi(x)-x\Phi^{'}(x)]/2x^2$ and $\Phi(x)$ is the error function \citep{vikhlinin01a}.

Following equation~\ref{eq:col_mfp} and the values of the thermodynamic quantities across the cold front (Table~\ref{tab:cf}), we derive $\lambda_e^{in} = 1.7$~kpc, $\lambda_e^{out} = 12.3$~kpc, $\lambda_e^{in \rightarrow out} = 14.7$~kpc and $\lambda_e^{out \rightarrow in} = 2.2$~kpc from the annular sector data. For the higher resolution ACB data, the quantities are calculated as $\lambda_e^{in} = 1.9$~kpc, $\lambda_e^{out} = 10.8$~kpc, $\lambda_e^{in \rightarrow out} = 11.6$~kpc and $\lambda_e^{out \rightarrow in} = 2.2$~kpc (Table~\ref{tab:cf}). If Coulomb diffusion is active in A3667 then we should expect the width of the cold front to be smeared up to several times that of $\lambda_e^{in \rightarrow out}$. Since the cold front width is {\bf $\Delta R = 24\pm12$~kpc}, the Coulomb diffusion has to be suppressed in A3667. 

{\it (ii) Thermal Conduction Across the Cold Front: } Further, we consider the effect of thermal conduction on the cold front. In order to estimate the characteristic time-scale of conduction across the width of the cold front, we follow the analysis for Abell 2142 as described in \citet{ettori00}. In the ICM, heat is conducted down the temperature gradient following \citep{sarazin88}:
\begin{equation}\label{eq:hflux}
{\bf Q}= \kappa \nabla{T_e}
\end{equation}
where ${\bf Q}$ is the heat flux, $\nabla{T_e} \approx \Delta T_e / \Delta R$ and $\kappa$ is the thermal conductivity given by \citep{cowie77}:
\begin{equation}\label{eq:kappa}
\kappa=1.31 n_e \lambda_e \left(\frac{k_B T_e}{m_e}\right)^{1/2}
\end{equation}
where $\lambda_e$ is the Coulomb mean free path and $m_e$ is the electron mass.

\citet{cowie77} showed that if the Coulomb mean free path is comparable to the scale of the thermal gradient ($\Delta R$) then the thermal conductivity can be suppressed by an order of magnitude from the Spitzer value due to the development of electrostatic fields. Moreover, the heat flux tends to saturate to a limiting value given by \citep{cowie77}:
\begin{equation}\label{eq:qsat}
{\bf Q_{sat}}=0.42 \left(\frac{k_B T_e}{\pi m_e}\right)^{1/2} n_e k_B T_e = 0.023 \left(\frac{k_B T_e}{10 keV} \right)^{3/2} \left(\frac{n_e}{10^{-3} cm^{-3}} \right)
\end{equation}
where the factor 0.42 is due to the reduction effect on the heat conduction caused by the development of the electrostatic field \citep{spitzer62}.
The maximum heat flux in a plasma is given by \citep{ettori00}:
\begin{equation}\label{eq:maxflux}
{\bf Q_{max}}=1.5 n_e k_B T_e \frac{dR}{d\tau}
\end{equation}
where $\delta \tau$ is the time-scale of conduction.

In the case of A3667, the error estimate on the width of the cold front is large {\bf ($\Delta R = 24\pm12$~kpc)}. Hence, we consider both the cases: a) the width of the cold front ($\Delta R$) comparable to the Coulomb mean free path $\lambda_e^{out}$ (Table~\ref{tab:cf}) and b) $\Delta R >> \lambda_e^{out}$. For the case a) $\Delta R \approx \lambda_e^{out}$ the heat flux will be saturated. Hence, to compute the characteristic time scale of conduction, we equate ${\bf Q_{sat}} = {\bf Q_{max}}$ [equations \ref{eq:qsat} and \ref{eq:maxflux}]. For case b) $\Delta R>>\lambda_e^{out}$, we compute the time scale of conduction by equating ${\bf Q} = {\bf Q_{max}}$ [equations \ref{eq:hflux} and \ref{eq:maxflux}]. The time-scales of the conduction for both the above cases are given by:
\begin{eqnarray}\label{eq:tau}
\delta \tau &=& 1.4 \pm 0.1 \times 10^6 yr  ~~~~\forall~~ \Delta R \approx \lambda_e^{out} \nonumber \\
\delta \tau &=& 8.2 \pm 1.0 \times 10^6 yr  ~~~~\forall~~ \Delta R >> \lambda_e^{out} 
\end{eqnarray}

\citet{roettiger99} concluded that the merger event in A3667 $\sim 10^9$~years old. Since the cold front is associated with the merger event in A3667, we compare the characteristic conduction time-scales (from equation~\ref{eq:tau}) with this age. The time required for thermal conduction with classical Spitzer conductivity (equation~\ref{eq:kappa}) to erase the thermal gradient across the cold front is $\lesssim 10^7$~years which is much smaller than the predicted age of the merger in A3667. Hence, we conclude that the thermal conduction should be suppressed by a factor of $\sim 100-700$ than the classical Spitzer value (equation \ref{eq:kappa}) in A3667. This is necessary to maintain the sharpness of the cold front seen in the X-ray observations. In comparison, \citet{ettori00} found the thermal conduction should be suppressed by a factor of $250-2500$ to account for the sharpness of the cold front in Abell 2142. 

The suppression of transport processes in the ICM is generally explained due to presence of magnetic fields \citep{carilli02b}. \citet{vikhlinin01a} derived the strength of the magnetic field to be $6 < B < 14~\mu G$ from the geometry of the cold front. For the above calculation of the isotropic thermal conduction (in absence of magnetic field) we have considered the width of the cold front ($\Delta R$) to be the characteristic length-scale of conduction. In presence of the magnetic field, the characteristic length-scale of heat exchange is reduced to the Larmor radius \citep{asai04,rosswog07}:
\begin{equation}\label{eq:larmor}
R_L=8.1\times 10^{-16} \left(\frac{B}{1\mu G} \right)^{-1} \left(\frac{T_x}{5~keV}\right)^{1/2} ~~ kpc 
\end{equation}
where $B \gtrsim 6 \mu G$ is the lower limit of magnetic field estimate from \citet{vikhlinin01a} and $T_x=T_{out}=7.48\pm0.13$ from Table~\ref{tab:cf}. The value of Larmor radius for A3667 comes out to be $R_L \lesssim 1.7\times 10^{-16}$~kpc. This result ($R_L << \Delta R$) suggests that in presence of magnetic field the thermal conduction will be suppressed and cannot erase the entire cold front (width of $\Delta R = 24\pm12$~kpc) in A3667 over $10^9$~years. 

\citet{asai05} used 3-dimensional MHD simulations with various conduction scenarios to reproduce the observed X-ray surface brightness of A3667 across the cold front. Their results clearly show that cold front in A3667 can last for $10^9$~years only in the presence of magnetic field and anisotropic heat conduction, where heat is conducted only in the direction of the magnetic field lines. Their results verified that isotropic thermal conduction destroys the cold front in $\lesssim 10^7$~years in absence of magnetic field. \citet{zuhone13} also carried out a series of MHD simulations in the core of galaxy clusters using anisotropic thermal conduction. Their results confirm that the sharp cold fronts observed in {\it Chandra} observations (e.g. A3667) cannot be reproduced with full Spitzer conduction. 

\subsection{Velocity of the Cool Gas}

%------------------------ Figure:- 16 -----------------------------
\begin{figure}[t!]
\centering
\epsfig{file=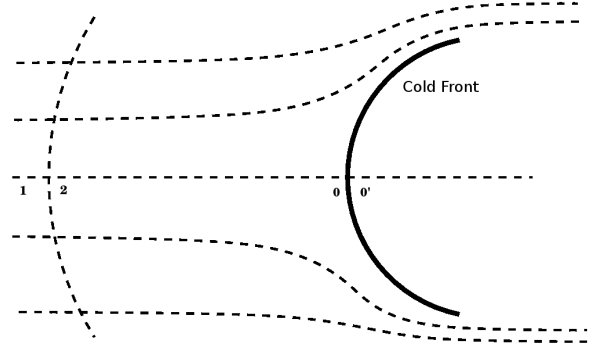,height=1.8truein}
\caption{Geometry of the flow of less dense ambient gas past a body of dense gas. The interface between this dense, colder gas and less dense hot gas is marked by the cold front in A3667 (solid curve between points 0 and 0'). Also shown the possible bow-shock region, dashed vertical region between points 1 and 2, as suggested by \citet{vikhlinin01b}.}
\label{fig:bow}
\end{figure}
%------------------------------------------------------------------
Here we investigate the existence of a possible bow-shock that was suggested in \citet{vikhlinin01b}. In order to do so, we need to first summarize the theory of supersonic flow past a blunt-edged body (as described in \citet{landau59}). 

In case of a supersonic flow past an arbitrary body, a shock wave is formed in front of the body (Figure~\ref{fig:bow}). If the body is not pointed then the shock front is detached from the body. In A3667 a similar scenario can be represented by the cold front. The cold and denser gas cloud inside the cold front can be treated as a body with a blunt-edge and the hot, less-dense ambient gas (away from the cold front) can be treated as a streamline flow. The point '0' in Figure~\ref{fig:bow} represents the 'stagnation point' where the relative velocity of the flow is zero with respect to the cold front. Along the streamline, the pressure at points '0' and '2' are related by equation 122.1 in \citet{landau59}:
\begin{equation}
P_0 = P_2 \left[ 1 + \frac{\gamma - 1}{2} M_2^2 \right]^{\frac{\gamma}{\gamma-1}}
\end{equation}
where $\gamma=5/3$ and $M_2$ is the Mach number at point 2. If the velocity of the body is supersonic then there is a bow shock between the points 2 and 1. Then the pressures $P_1$ and $P_2$ will be related by a shock jump condition. However, if the velocity of the body is not supersonic then a shock is not created and $P_1=P_2$. Hence, if the velocity of the body is sub-sonic $P_0$ and $P_1$ are related by:
\begin{equation}\label{eq:pratio}
P_0 = P_1 \left[ 1 + \frac{\gamma - 1}{2} M_1^2 \right]^{\frac{\gamma}{\gamma-1}}
\end{equation}

\citet{vikhlinin01b} derived the pressure jump between the points '0' and '1' to find $P_0/P_1=2.09 \pm 0.48$ which corresponds to a Mach number of $\mathcal{M}=1.0 \pm 0.2$. They also detected a jump in X-ray surface brightness and temperature about $350$~kpc in front of the cold front. They proposed this to be the location of a possible bow-shock. Such a bow-shock is also predicted from the theory (discussed above) only if the flow is supersonic. In our analysis, we obtained the pressure just outside the cold front to be $P_0=10.9 \pm 0.5$ and about $400$~kpc away from the cold front to be $P_1=8.6 \pm 0.9$. Both $P_0$ and $P_1$ are in units of $1.6\times10^{-13}$~Pa. Hence, the resultant pressure jump is $P_0/P_1 = 1.3 \pm 0.2$. Using equation~\ref{eq:pratio}, we obtain the value of the Mach number $M_1=0.6 \pm 0.2$. Therefore the velocity of the dense gas cloud inside the cold front can be derived as\citep{vikhlinin01b}:
\begin{equation}\label{eq:vel}
v=M_1 c_1 = M_1 \left(\frac{\gamma T_1}{m_p\mu}\right)^{1/2}
\end{equation}
where $\mu=0.6$ is the mean molecular weight of the intracluster plasma and $T_1 = 8.5 \pm 0.9$~keV (Figure~\ref{fig:cf_pie}e). Therefore, we get the velocity of the cloud to be $v = 900 \pm 330~km~s^{-1}$. \citet{vikhlinin01b} estimated the velocity of cold, dense gas cloud to be $1430 \pm 290~km~s^{-1}$. In our analysis, we found the velocity of the cold front to be sub-sonic with respect to the ambient which suggests that we should not see a bow shock within $\sim 400$~kpc in front of the cold front. In Figures~\ref{fig:cf_pie} and \ref{fig:cf_acb} we have shown the profiles of thermodynamic quantities up to $\sim 400$~kpc away from the cold front. Both these Figures do not show any hint of a jump in X-ray surface brightness or temperature in the proposed location of the bow-shock \citep{vikhlinin01b}. Moreover, the Mach number maps in Figure~\ref{fig:Amach} also do not show any signature of a bow shock near the proposed location. This can be due to the fact that the proposed location of this bow shock falls very near the edge of the field-of-view in the current {\it Chandra} observations where we do not have sufficient sensitivity to detect small jumps in temperature or the X-ray surface brightness. Hence, it is not possible to infer anything about the existence of this possible bow shock from the current X-ray data as suggested in \citet{vikhlinin01b}. On the contrary, the derived value of $M_1=0.6 \pm 0.2$ suggests that there should not be any bow shock at locations further away from the cold front. 

\section{Conclusion}
\label{sec:conc}

In this section, we summarize the major results and conclusions from this paper. Here, we have analyzed the extensive {\it Chandra} archival data of A3667 and have produced new high resolution, large field-of-view temperature maps for the cluster (Figures~\ref{fig:Atemp_wvt} and ~\ref{fig:Atemp_acb}). We have compared two temperature map making methods (WVT and ACB) and discussed in detail their relative advantages. The advantage of the WVT method is that it produces spatially non-overlapping regions which are statistically independent. We can propagate errors from WVT temperature maps to evaluate the errors in the resulting Mach number map. However, the major limitation of the WVT map is that it can miss some temperature structure near low SNR regions. The ACB method complements the WVT method in this aspect. Due to a different binning kernel, the ACB method can detect temperature structures at scales below the individual WVT regions. However, the adjacent ACB regions are highly correlated. Hence, it is impossible to derive statistically independent error estimates for temperatures between neighboring pixels. Therefore, combined interpretations of both the WVT and ACB results are required to extract relevant temperature structure from any galaxy cluster X-ray data. Computationally, ACB is an expensive process. Hence, we advise to run the WVT temperature map first in order to find the appropriate SNR value and then proceed with the ACB method. 

Since the effective exposure time of all the combined {\it Chandra} observations is very high ($\sim 447$~ksec), the correlation length of the ACB circles is very small as compared to the field-of-view in case of A3667 (Figure~\ref{fig:Atemp_acb}(b)). Due to high exposure time, the WVT temperature map (Figure~\ref{fig:Atemp_wvt}) was able to extract major structures in temperature as in the ACB temperature map (Figure~\ref{fig:Atemp_acb}). The ACB method becomes more relevant for observations with limited exposure time. 

In this paper, we proposed a new shock-finder which can be used on the observed 2-dimensional X-ray data of galaxy clusters. Here, we have applied this shock-finder on the A3667 X-ray data as well as the projected X-ray surface brightness and temperature maps from a simulated cluster from cosmological MHD AMR simulations. We have successfully validated our observational shock-finder results on the 2-dimensional projected data of a simulated MHD cluster with the results from the original 3-dimensional shock-finder used on the 3-dimensional data from cosmological MHD AMR simulations of the same cluster. Moreover, we combined the shock statistics of A3667 X-ray/radio data and compared them with the shocks from the original MHD cluster. We concluded that a combination of observations at X-ray and radio wavelengths is crucial to analyze the shocks in a galaxy cluster (like A3667) accurately. This is one of key results from our work. Moreover, the striking agreement between the shock statistics between the observations and simulations (Figure~\ref{fig:machhist}) suggests that the shocks in a real observed galaxy cluster like A3667 can be reproduced in the current state-of-the-art AMR simulations [e.g. \citet{skillman13}] due to accurate modeling of the relevant physical processes in the ICM. Thus, we can gain more insights from the 3-dimensional MHD AMR simulations while analyzing the galaxy clusters observed at X-ray/radio wavelengths. With the growing synergy between observations and simulations, the simulations hold the key to predict observables with the increased sensitivity of upcoming or future telescopes both at radio (e.g. LOFAR, MWA, EVLA,SKA, LRA) and X-ray (e.g. Astro-H) wavelengths.

With the detection of the radio bridge and overlapping central X-ray emission connecting with the emission near the relic \citep{carretti12}, future {\it Chandra} observations near the radio relic will help us understand the X-ray activity in the region. The limited field-of-view of the current {\it Chandra} archival data on A3667 did not have any coverage near the North-West radio relic. The only data available is from XMM-Newton \citep{finoguenov10}. A higher resolution {\it Chandra} observation will help us to get more accurate shock statistics near the radio-relic and compare them with the Mach number distribution derived from the radio data. Since a shock front is comprised of a distribution of Mach numbers \citep{skillman13}, a comparison of the radio Mach number distribution with a single Mach number value from XMM analysis is not sufficient to check the agreement between the radio and X-ray Mach numbers. \citet{ogrean13} summarizes the agreement between radio and X-ray Mach numbers in clusters Abell 754, CIZA J2242.8+5301, Abell 3376 and Abell 521. This is not true in the case of 1RXS J0603.3+4214 (Toothbrush Relic), where the X-ray Mach number is lower than the Radio Mach number by a factor of $\gtrsim 1.7$ which can be due to the fact that the synchrotron emission is more sensitive to high Mach numbers \citep{hoeft07}. With high resolution of the {\it Chandra}, it will be possible to investigate the agreement between radio and X-ray Mach numbers for A3667 near the radio-relics. 

We have also investigated the profiles of the thermodynamic properties across the cold front in A3667. In order to derive the deprojected electron density across the cold front, we have used the spherical deprojection technique. This is not the suitable for most of the cluster geometry. For A3667, most of the merger activities are in the plane of the sky \citep{carretti13}, hence the assumption about the spherical symmetry may not have much effect on estimating the real electron density. Previously, \citet{donahue06} also found very little effect of the symmetry assumption on the final entropy profiles.

From the results of the thermodynamic properties across the cold front, we constrained the width of the cold front to $\Delta R = 24\pm12$~kpc. The only previous study by \citet{vikhlinin01b} that constrained the width of the cold front in A3667 was based on a short {\it Chandra} observation ($49$~ksec). They constrained the width to $\sim 2$~kpc only from the X-ray surface brightness profile as the temperatures were extracted over much larger bin sizes $\gtrsim 40$~kpc. Our current paper, is the first study that constrains the width of the cold front in A3667 from both the temperature and surface brightness data. This have been possible due to the combined exposure time of eight {\it Chandra} observations. We have further investigated the effect of transport processes like thermal conduction and Coulomb diffusion on the stability of the cold front. Our analysis suggest that the thermal conduction across the cold front should be suppressed by a factor of $\sim 100-700$ than the classical Spitzer value. This will allow the cold front to last over $10^9$~years, the estimated current age of the merger in A3667 \citep{roettiger99}. 

Finally, we find the velocity of the cold front with respect to the ambient is sub-sonic. This in turn suggests that it is very unlikely to find a bow shock upstream ($\sim 350$~kpc) of the cold front. Moreover, the profiles of density, temperature, pressure and entropy do not show any hint of a shock at this location as suggested by \citet{vikhlinin01b}. Since this location falls very near the edge of the field-of-view of the current {\it Chandra} observations, we do not have sufficient sensitivity to infer anything about the existence of this bow-shock. Future observations are required to improve the sensitivity at this location and investigate the possibility of a bow-shock.
Our analysis suggests that cold fronts can be utilized to gauge mergers in galaxy clusters, provided we have sufficient sensitivity in the X-ray observations. 

\vskip 0.3in

We would like to thank Matt Owers for providing the exact locations of the optical substructures (shown in Figure~\ref{fig:rgb}) and Huub R\"{o}ttgering for providing the FITS images of A3667 from ATCA observations. We would also like to thank Hao Xu for the original MHD cluster from \citet{xu11} that has been used in \citet{skillman13} and further used to compare with the A3667 results in this work. AD has been supported by NASA Postdoctoral Fellowship Program through NASA Lunar Science Institute. This work was funded by National Science Foundation (NSF) grant AST 1106437 to J.O.B. This work utilized the Janus supercomputer, which is supported by the NSF (award number CNS-0821794) and the University of Colorado, Boulder. The Janus supercomputer is a joint effort of the University of Colorado Boulder, the University of Colorado Denver, and the National Center for Atmospheric Research. The LUNAR Consortium (http://lunar.colorado.edu), headquartered at the University of Colorado, is funded by the NASA Lunar Science Institute (via cooperative Agreement NNA09DB30A), and partially supported this research.

\bibliographystyle{apj}
%\bibliography{ms}

\end{document}